\documentclass[apj,onecolumn,dvipdfm]{emulateapj}
\usepackage{graphicx,epsf}
\newcommand{\ltsim}{\protect\raisebox{-0.5ex}{$\:\stackrel{\textstyle <}{\sim}\:$}}
\newcommand{\gtsim}{\protect\raisebox{-0.5ex}{$\:\stackrel{\textstyle >}{\sim}\:$}}

\shorttitle{Non-Ideal RMHD Simulations of Protostellar Collapse}
\shortauthors{Tomida et al.}
\begin{document}
\title{Radiation Magnetohydrodynamic Simulations of Protostellar Collapse: Non-Ideal Magnetohydrodynamic Effects and Early Formation of Circumstellar Disks}

\author{Kengo Tomida\altaffilmark{1,2}, Satoshi Okuzumi\altaffilmark{3}, and Masahiro N. Machida\altaffilmark{4}}
\altaffiltext{1}{Department of Astrophysical Sciences, Princeton University, Princeton, NJ 08544, USA; \mbox{tomida@astro.princeton.edu}}
\altaffiltext{2}{Department of Physics, The University of Tokyo, Tokyo, 113-0033, Japan}
\altaffiltext{3}{Department of Earth and Planetary Sciences, Tokyo Institute of Technology, Meguro-ku, Tokyo, 152-8551, Japan; \mbox{okuzumi@geo.titech.ac.jp}}
\altaffiltext{4}{Department of Earth and Planetary Sciences, Faculty of Sciences, Kyushu University, Hakozaki, Higashi-ku, Fukuoka 812-8581, Japan; \mbox{machida.masahiro.018@m.kyushu-u.ac.jp}}

\begin{abstract}
The transport of angular momentum by magnetic fields is a crucial physical process in formation and evolution of stars and disks. Because the ionization degree in star forming clouds is extremely low, non-ideal magnetohydrodynamic (MHD) effects such as ambipolar diffusion and Ohmic dissipation work strongly during protostellar collapse. These effects have significant impacts in the early phase of star formation as they redistribute magnetic flux and suppress angular momentum transport by magnetic fields. We perform three-dimensional nested-grid radiation magnetohydrodynamic (RMHD) simulations including Ohmic dissipation and ambipolar diffusion. Without these effects, magnetic fields transport angular momentum so efficiently that no rotationally supported disk is formed even after the second collapse. Ohmic dissipation works only in a relatively high density region within the first core and suppresses angular momentum transport, enabling formation of a very small rotationally supported disk after the second collapse. With both Ohmic dissipation and ambipolar diffusion, these effects work effectively in almost the entire region within the first core and significant magnetic flux loss occurs. As a result, a rotationally supported disk is formed even before a protostellar core forms. The size of the disk is still small, about 5 AU at the end of the first core phase, but this disk will grow later as gas accretion continues. Thus the non-ideal MHD effects can resolve the so-called magnetic braking catastrophe while maintaining the disk size small in the early phase, which is implied from recent interferometric observations.
\end{abstract}

\keywords{stars: formation --- ISM: clouds --- ISM: jets and outflows --- radiative transfer --- magnetohydrodynamics}

\section{Introduction}
Circumstellar disks are supposed to form as natural by-products of star formation processes due to large angular momenta in natal molecular cloud cores. Although extensive observational and theoretical studies have been done, theories of disk formation are not complete yet. A disk plays crucial roles in star formation and evolution since its presence affects the evolution of the central star and its feedback. Moreover, circumstellar disks, or protoplanetary disks, are very important not only in the context of star formation but also of planet formation because they are the sites of planet formation. Therefore, it is of crucial importance to construct a consistent model of circumstellar disk formation in the context of star formation.

The key to understand disk formation is angular momentum redistribution during protostellar collapse. Gravitational torque is important when magnetic fields are weak \citep{bate98,saigo08,bate10,tomida10b,bate11}, but typically magnetic fields observed in molecular clouds \citep[][and references therein]{crutcher,hlipp6} are strong enough to remove angular momenta efficiently and overcome the centrifugal barrier \citep{mp79,mp80,bm94,bm95a,bm95b,tmsk98,tmsk00,tmsk02,mcd05a,mcd05b,com10,tomida10a,mim11b,bate14}. However, this magnetic braking is in fact too effective and strongly suppresses formation of rotationally supported disks and binary/multiples at least in the early phase of star formation and with the ideal MHD approximation, which is called the magnetic braking catastrophe \citep[e.g.][]{ms56,ml08,allen03,li11,zlipp6} or fragmentation crisis \citep{ht08}.

Obviously this problem cannot be real because many protoplanetary disks and extra-solar planets have been observed, and because the fraction of binaries or multiples is known to be high \citep[e.g.][]{rag10,dk13}. Therefore, there must be something missing in those early simulations which were highly idealized and simplified. This problem has been actively discussed recently and many solutions have been proposed. For example, non-ideal MHD effects such as Ohmic dissipation and ambipolar diffusion can redistribute magnetic flux and suppress angular momentum transport \citep{db10,mm11,mim11b,dapp12,mh13,tomida13}. Mis-alignment between magnetic fields and the rotational axis can reduce the efficiency of magnetic braking and enable disk formation when magnetic fields are not too strong \citep{joos12,kch13}. Or simply, a circumstellar disk can form later when the surrounding envelope almost disappears and magnetic braking becomes inefficient \citep{ml09,mim11,mh13}. Turbulence that ubiquitously exists in star forming clouds is another promising mechanism, but there are various interpretations how it circumvents the magnetic braking catastrophe; diffusion or reconnection enhanced by turbulence can redistribute magnetic fields \citep{lv99,sl12,sl13}, incoherent magnetic fields are less efficient to transport angular momentum \citep{sei13,sei14}, it can naturally induce mis-alignment between magnetic fields and rotation \citep{joos13}, or simply turbulence carries additional angular momentum locally. Given such many solutions proposed, the magnetic braking catastrophe is not so catastrophic as it sounds, but it still remains as quantitative questions; when and how is a disk formed, and how large (massive) is it?

It is observationally well known that star forming clouds are strongly magnetized. Magnetic fields in dense molecular cloud cores are slightly weaker than or comparable to the critical field strength required to support those cloud cores \citep{crutcher,hli13}. Observations of circumstellar disks around young stellar objects also have been actively carried out. \citet{maury} performed interferometric observations toward Class-0 sources, and compared them with numerical simulations with and without magnetic fields. They did not detect largely-extended circumstellar disks, suggesting that the effects of magnetic fields are significant in formation and early evolution of circumstellar disks. \citet{tobin12,tobin13} \citep[see also][]{sakai14} claimed that they found a large circumstellar disk with a radius of $\sim 120\, {\rm AU}$ around a very young Class-0 object L1527 IRS, one of the best-studied examples. While this result was striking, \citet{ohashi} observed the same object with higher resolution and sensitivity using the Atacama Large Millimeter/submillimeter Array (ALMA), and concluded that the disk radius should be about or smaller than $60\, {\rm AU}$. Clearly more detailed and extensive observations are needed, but overall, these observations imply that young circumstellar disks should be small, while they could form in the early phase of star formation and should grow later. 

Thus magnetic fields play crucial roles in star formation, and therefore their effects must be carefully clarified and quantified. Because the ionization degree in star forming clouds are extremely low, redistribution of magnetic fields by non-ideal MHD effects is naturally expected and has a significant impact on angular momentum transport \citep{wk93,fm93,bm94,cm94,dm01,nkn02,tm07,ml09,kunz09,kunz10,li11,dapp12}. Non-ideal MHD effects are also important in the context of the magnetic flux problem, which means that typical magnetic flux in initial molecular cloud cores is far larger than that in formed stars by orders of magnitude and therefore substantial flux redistribution is required \citep[e.g.][]{pm83,bra12}. In this work, we investigate the non-ideal MHD effects and disk formation during protostellar collapse. For this purpose, we perform three-dimensional RMHD simulations of protostellar collapse including non-ideal MHD effects. In \citet{tomida13} (hereafter Paper I), we found that Ohmic dissipation can redistribute magnetic flux from the dense region and enable formation of rotationally supported disks around protostellar cores. In this paper, we extend our simulations to include ambipolar diffusion, which is more efficient in the lower density region. This paper is organized as follows. The equations and numerical methods are explained in Section 2, and the models used in the simulations are described in Section 3. The results of the numerical simulations are presented in Section 4. Section 5 is devoted to discussion and conclusions are summarized in Section 6. In the Appendix, we describe the newly implemented solver for the non-ideal MHD effects and demonstrate its validity.

\section{Methods}
\subsection{Basic Equations}
In order to study protostellar collapse, many physical processes must be properly taken into account, such as MHD, self-gravity, radiation transfer, chemistry and non-ideal MHD effects. In Paper I, we only considered Ohmic dissipation as non-ideal MHD effects. In this work, we extend our three-dimensional (3D) nested-grid simulations to include ambipolar diffusion, which is caused by decoupling between neutral and charged particles. We adopt the single fluid approach assuming strong coupling \citep{ml95,dp08,masson}. Note that the Hall effect, another possibly important non-ideal MHD effect \citep{li11,bw12}, is not included in this work. The governing equations we use are as follows:
\begin{eqnarray}
\frac{\partial\rho}{\partial t}+\nabla\cdot(\rho\mathbf{v})=0,\label{mc}\\
\frac{\partial\rho\mathbf{v}}{\partial t}+\nabla\cdot\left[\rho\mathbf{v}\otimes\mathbf{v}+\left(p+\frac{1}{2}|\mathbf{B}|^2\right)\mathbb{I}-\mathbf{B}\otimes\mathbf{B}\right]=-\rho\nabla\Phi+\frac{\sigma_R}{c}\mathbf{F}_r,\\
\frac{\partial\mathbf{B}}{\partial t}-\nabla\times\left(\mathbf{v}\times\mathbf{B}-\eta_{\rm O}\mathbf{J}-\frac{\eta_{\rm A}}{|\mathbf{B}|^2}\mathbf{B}\times\mathbf{F}\right)=0,\label{induction}\\
\frac{\partial e}{\partial t}+\nabla\cdot\left[\left(e+p+\frac{1}{2}|\mathbf{B}|^2\right)\mathbf{v}-\mathbf{B}(\mathbf{v}\cdot\mathbf{B})+\eta_{\rm O}\mathbf{F}+\frac{\eta_{\rm A}}{|\mathbf{B}|^2}(\mathbf{B}\times\mathbf{F})\times\mathbf{B}\right]=\nonumber\\
-\rho\mathbf{v}\cdot\nabla\Phi-c\sigma_P(aT_g^4-E_r)+\frac{\sigma_R}{c}\mathbf{F}_r\cdot\mathbf{v},\label{energy}\\
\mathbf{J}\equiv \nabla\times \mathbf{B},\hspace{2em} \mathbf{F}\equiv \mathbf{J}\times \mathbf{B},\nonumber\\
\nabla\cdot\mathbf{B}=0,\label{sole}\\
\nabla^2\Phi=4\pi G\rho,\\
\frac{\partial E_r}{\partial t}+\nabla\cdot(\mathbf{v}E_r)+\nabla\cdot\mathbf{F}_r+\mathbb{P}_r:\nabla\mathbf{v}=c\sigma_P(a_r T_g^4 -E_r),\\
\mathbf{F}_r=\frac{c\lambda}{\sigma_R}\nabla E_r,\hspace{1em}
\lambda(R)=\frac{2+R}{6+2R+R^2},\hspace{1em}
R=\frac{|\nabla E_r|}{\sigma_R E_r},\nonumber\\
\mathbb{P}_r=\mathbb{D}E_r,\hspace{1em}
\mathbb{D}=\frac{1-\chi}{2}\mathbb{I}+\frac{3\chi-1}{2}\mathbf{n}\otimes \mathbf{n},\hspace{1em}
\chi=\lambda+\lambda^2R^2,\hspace{1em}
\mathbf{n}=\frac{\nabla E_r}{|\nabla E_r|}\nonumber,\label{end}
\end{eqnarray}
where $\rho, \mathbf{v}, \mathbf{B}, p, T_g, e, E_r, \mathbf{F}_r, \mathbb{P}_r, \Phi$ are the gas density, gas velocity, magnetic flux density, gas pressure, gas temperature, total gas energy, radiation energy, radiation flux and radiation pressure tensor, respectively. $c=2.99792 \times 10^{10} \, {\rm cm\,s^{-1}}$ is the speed of light, $G= 6.673 \times 10^{-8} \, {\rm cm^3\,g^{-1}\,s^{-2}}$ is the gravitational constant, and $a_r=7.5657 \times 10^{-15}{\rm erg \, cm^{-3} \, K^{-4}}$ is the radiation density constant. From top to bottom, these equations represent conservation of mass, the equation of motion, the induction equation including Ohmic dissipation and ambipolar diffusion, the gas energy equation, the solenoidal constraint, the Poisson's equation of gravity, and the radiation transfer equations. We adopt the gray flux limited diffusion approximation for radiation transfer \citep{lp81,lev84}. The tabulated Ohmic resistivity $\eta_{\rm O}$ and ambipolar diffusion rate $\eta_{\rm A}$ are calculated using a chemical network (see \ref{srate}). For the Rosseland and Planck mean opacities, $\sigma_R$ and $\sigma_P$, we combine three tables; \citet{semenov}, \citet{op94} and \citet{fer05}.

\subsection{Overview of the Numerical Scheme}
In order to achieve the huge dynamic range to resolve star formation processes from the molecular cloud core scale down to the stellar core scale, we use the 3D self-similar nested-grid technique \citep{yk95,zy97,mcd05a,mcd05b,tomida13}. We solve these equations using the simple operator-splitting technique, using the same time steps over the whole nested-grid hierarchy (the so-called shared time stepping). Finer levels are generated at the center of the parent level so that the local Jeans length is resolved at least with 16 cells \citep{trlv97,com08,joos12}. We stop the simulations when this condition cannot be satisfied, as unphysical fragmentation may occur. Each level consists of $64^3$ cells, and 14 levels are created by the end of the first core phase. Typical resolution in the first core (3-5 AU from the center of the cloud, which is covered with level l=12) is $\Delta x \sim 0.14 \, {\rm AU}$ or higher. The MHD part is solved using the HLLD approximate Riemann solver \citep{miyoshi} and the mixed divergence cleaning method \citep{dedner}. The multigrid solver \citep{mh03} is used for self-gravity. The time step for the MHD and radiation transfer parts is determined by the Courant-Friedrich-Levy (CFL) condition for the MHD part, and the radiation subsystem is solved implicitly. The implementation of newly-introduced non-ideal MHD effects is described in the Appendix. Detailed descriptions of other parts including microphysical processes are given in Paper I.

Only in the model with ambipolar diffusion, we additionally introduce a density floor where magnetic fields are dominantly strong so that the time step does not become too small. When the gas density is so low that the plasma beta $\beta \equiv P_{\rm gas}/P_{\rm mag}$ gets below $10^{-3}$, the gas density is increased so that $\beta$ becomes $10^{-3}$ while the gas temperature and velocity are unchanged. This floor works only in the very low density region in the infalling envelope just above the first core. We monitor the mass added by this density floor and confirm it is very small, about $5 \times 10^{-5}\, M_\odot$ in total, or less than 0.1\% of the first core mass. Even with this density floor, the time step is limited by the Alfv\'{e}n speed in the low density region above the first core in the model with ambipolar diffusion, and as a result an excessive number of time steps are required.

It is important to solve the energy equations including the effects of radiation transfer, although radiation does not drastically affect the evolution in the early phase of star formation compared to the previous simulations using the barotropic approximation \citep[e.g.][]{bate98,saigo08,mim08,mtmi08}. Radiation heating and cooling are only crudely treated in the barotropic approximation, and heating by the non-ideal MHD effects is completely neglected. In reality, the thermal evolution varies depending on many factors, but the barotropic approximation simply assumes that the gas temperature is a function of the local gas density. Therefore, it cannot reproduce the realistic evolution and may affect the results artificially. For example, the stability of the first core depends on the gas temperature especially when it is gravitationally unstable \citep{sw09}. Also, the heating by the non-ideal MHD effects affects the structure of the first core (see Paper I), and the rates of the non-ideal MHD effects are sensitive to the gas temperature. Thus it is crucial to treat both dynamics and thermodynamics consistently.

\subsection{Microphysics}
\subsubsection{Update on the EOS}
We update the equation of state from Paper I. Now the partition function for vibration of molecular hydrogen is
\begin{eqnarray}
Z_{\rm vib,H_2} = \frac{1}{1-\exp{\left(\frac{\theta_{\rm vib}}{T}\right)}},
\end{eqnarray}
where $\theta_{\rm vib}\sim 6330\, {\rm K}$ is the excitation temperature of vibration of molecular hydrogen\footnote{For example, NIST Computational Chemistry Comparison and Benchmark Database, NIST Standard Reference Database Number 101, Release 16a, http://cccbdb.nist.gov/}. This formula is better than the previous one used in Paper I because it does not diverge in the low temperature limit. However, this change has very minor impacts on the results of hydrodynamic simulations because contribution from the vibrational degrees of freedom becomes significant only in a high temperature, and most of hydrogen molecules are already dissociated there. Other parameters are unchanged including the ortho:para ratio of molecular hydrogen, which is 3:1.

\subsubsection{Ohmic Resistivity and Ambipolar Diffusion Rate} \label{srate}
The Ohmic resistivity and ambipolar diffusion rate are calculated based on the method described in \citet{nkn02} and \citet{okz09}. This model solves a chemical reaction network of ${\rm H_3^+, HCO^+, He^+, Mg^+, C^+, H^+}$ and dust grains, and derive equilibrium solutions. In addition, we consider thermal ionization of potassium (K), which produces a sudden increase of the ionization degree and sharp cut-off in the rates of the non-ideal MHD effects around $T=650\, {\rm K}$. We adopt the reaction rates from the UMIST RATE 2012 database \citep{umist}. Dust grains can be charged from -14 to +14. We assume that the dust to gas ratio is 0.01, the dust density is $2 \,{\rm g \,cm^{-3}}$, and the size of the dust particles is $0.1\, {\rm \mu m}$. The Ohmic resistivity in this work is higher than that in Paper I in the low density region, by about a factor of 3 in most regions except for the sharp increase around $\rho\sim 10^{-14} \, {\rm g\, cm^{-3}}$ (see Figure~\ref{rate} below). In Paper I, this rise occurs around a higher density, $\rho\sim 10^{-13} \, {\rm g\, cm^{-3}}$. These differences result in larger angular momenta remaining in the first cores, but the results are still qualitatively similar.

We simply adopt a constant ionization rate $1.0\times 10^{-17} \, {\rm s^{-1}}$, assuming cosmic rays are the dominant ionization source and neglecting their shielding. In dense regions where cosmic rays cannot penetrate, this assumption overestimates the ionization rate and hence underestimates the diffusion rates. The ionization rate in such regions is generally determined by decay of radioactive elements. At least for our solar system, there is meteoritic evidence that short-lived radionuclides such as ${\rm ^{26}Al}$ existed when the solar nebula formed \citep{adams10}. As discussed in Paper I, these short-lived nuclides yield an ionization rate of $0.7 - 1.0 \times 10^{-18} \, {\rm s^{-1}}$ \citep{un09}. In this case, neglecting cosmic-ray shielding would have only a moderate impact. By contrast, the ionization rate in dense regions can be extremely low if only long-lived radio active nuclides such as ${\rm ^{40}K}$ exist \citep{un09,kunz09,kunz10,dapp12}. Because the lifetime of ${\rm ^{26}Al}$ is short and is comparable to the time scale of star formation, its abundance would vary in each system. Therefore this is an environmental parameter rather than an uncertainty. In this work, we aim to demonstrate that disk formation is possible even with the conservative case with the high ionization rate (i.e. the resulting non-ideal MHD effects are weak) as the first step.

Another significant simplification in our model is that dust particles have a single population of $0.1\, {\rm \mu m}$, neglecting their distribution and evolution. While dust grains can grow in collapsing clouds, they also can be spattered at shocks. On the other hand, dust grains evaporate at high temperatures. Although the effects of dust evaporation are taken into account in the opacities \citep{semenov}, they are not included in our rates. Specifically, water ices evaporate around $T\sim 150\, {\rm K}$, while silicates evaporate around $T\sim \, 1,400{\rm K}$. The latter has no significant impact on the rates because thermal ionization of potassium occurs below that temperature, and magnetic fields and gas are well coupled there. The evaporation of water ices, however, can be important but its effect strongly depends on dust properties such as composition and structures, which are highly uncertain.

Thus the rates of the non-ideal MHD effects have large uncertainties, and they would vary in different environments. Therefore, despite the detailed calculation of the rates, our simulations should be considered as an experiment with a typical model, and broad parameter surveys must be performed in the future.

\section{Models}
\begin{table*}[tbp] 
\begin{center}
\caption{Summary of the initial model parameters and results}
\begin{tabular}{c||ccccc|ccccc}
Model & $\Omega \, (\times 10^{-14}\, {\rm s^{-1}})$ & $B_0 \, ({\rm \mu G})$ & $\mu_0$  & OD & AD & $\tau_{\rm FC} ({\rm yrs})$& $M_{\rm FC}$ & $M_{\rm FC}/\Phi_{\rm FC} ({\rm cgs})$ & RSD Formation & $R_{\rm disk} ({\rm AU})$\\
\hline
{\it I}  & 2.4 & 20 & 3.8 & N & N &   770 &  0.03 &  2,500 &             Not Formed & --\\
{\it O}  & 2.4 & 20 & 3.8 & Y & N & 1,050 &  0.04 &  7,200 &  After Second Collapse & 1\\
{\it OA} & 2.4 & 20 & 3.8 & Y & Y & 2,750 & 0.075 & 37,000 & Before Second Collapse & 5
\end{tabular}
\end{center}
{{\bf Notes.} The first five columns are the initial model parameters: the angular velocity, the magnetic field strength, the normalized mass-to-flux ratio, whether Ohmic dissipation and ambipolar diffusion are introduced. Other parameters are common: $M=1\,M_\odot$, $R\sim 8800\, {\rm AU}$, $\rho_c=1.2\times 10^{-18}\, {\rm g\, cm^{-3}}$, $T_0=10\,{\rm K}$ and $A_2=0.1$. The last five columns indicate the results of the simulations: the first core lifetime, the first core mass, the first core mass-to-flux ratio (not normalized), remarks on formation of rotationally supported disks, and the disk radius at the ends of the simulations. See the text for details.}
\label{table:models}
\end{table*}

We adopt the same initial conditions as in Paper I. An unstabilized Bonnor-Ebert \citep[hereafter BE,][]{bonnor,ebert} sphere with m=2 perturbation is used as the initial density profile:
\begin{eqnarray}
\rho(r)=\rho_{\rm BE}(r)(1+A_0)\left[1+A_2\frac{r^2}{R^2}\cos(2\phi)\right],
\end{eqnarray}
where $\rho_{\rm BE}(r)$ is the the critical BE sphere with $\rho_c=10^{-18}\,{\rm g \, cm^{-3}}$ and $T=10\, {\rm K}$, $R$ the radius of the critical BE sphere, $A_0$ the amplitude of the initial density enhancement, $A_2$ the amplitude of regularized $m=2$ perturbation, respectively. We adopt $A_0=0.2$ and $A_2=0.1$. The mass and radius of this cloud are $M\sim 1M_\odot$ and $R\sim 4.25\times10^{-2}\,{\rm pc}\sim 8800\,{\rm AU}$. The initial free fall time at the center of the cloud is $t_{\rm ff}\sim 6.08\times 10^4 \,{\rm years}$. 

We initialize the cloud with solid-body rotation and uniform magnetic fields both aligned to $z$-axis. The angular rotation speed is $\Omega=2.4 \times 10^{-14}\, {\rm s^{-1}}$ or $\Omega t_{\rm ff}=0.046$. The magnetic field strength is $B_z=20\,{\rm \mu G}$ and the corresponding mass-to-flux ratio normalized by the critical value of stability for a uniform sphere is $\mu_0\equiv \frac{M/\Phi}{(M/\Phi)_{\rm crit}}\sim 3.8$ where $\Phi=\pi R^2 B_0$ and $(M/\Phi)_{\rm crit}=\frac{0.53}{3\pi}\left(\frac{5}{G}\right)^{1/2}$ \citep{ms76}. These parameters are the same to those in the fast rotating models {\it RF} and {\it IF} in Paper I.

The boundary conditions are also similar to those in Paper I. All the variables except magnetic fields outside the initial BE sphere are fixed to their initial values. Periodic boundary conditions are adopted for magnetic fields in order to avoid the divergence error. Because the gas outside the BE sphere is fixed, the magnetic fields are also anchored to the ambient gas. The boundary condition for self-gravity is set by the multi-pole expansion \citep{mh03}. These model setups are chosen in order to model isolated dense molecular cloud cores.

We calculate three models; Model {\it I} is the ideal MHD approximation, Model {\it O} includes Ohmic dissipation only, and Model {\it OA} includes both Ohmic dissipation and ambipolar diffusion. The model parameters are summarized in Table~1. While Model {\it I} evolves without any trouble until the central temperature exceeds $T_c \sim 80,000\, {\rm K}$, the other two models are more troublesome. As a result of suppression of angular momentum transport, rotationally supported disks are formed in these two models. These disks become gravitationally unstable after the second collapse in Model {\it O} and during the second collapse in Model {\it OA}. Because of the limitation of the nested-grid code, we only can refine the center of the computational domain, and therefore we cannot follow further evolution when fragmentation occurs outside the finest grid. Thus we stop the calculations around $T_c \sim 30,000 \, {\rm K}$ in Model {\it O} and $T_c \sim 2,100 \, {\rm K}$ in {\it OA}. For this reason, we focus on the first core phase in this work. Larger grids or more flexible adaptive mesh refinement (AMR) is required for further studies after fragmentation occurs.

\section{Results}

\begin{figure}[t]
\begin{center}
\scalebox{0.85}{\includegraphics{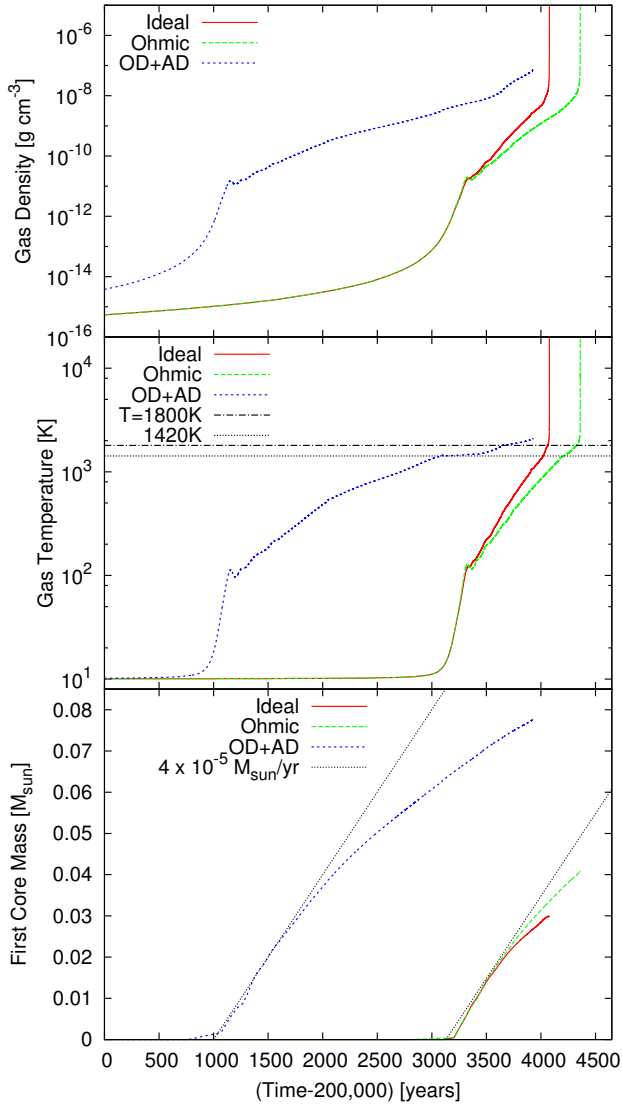}}
\caption{Time evolution of the gas density (top) and temperature (middle) at the center of the clouds, and the first core masses (bottom). The temperatures where the dust evaporation completes and where the second collapse begins are also shown with the gray dotted and dash-dotted lines. Note that these temperatures depend on the gas density, but the dependency is weak and the densities where these events happen in these models are close enough (see Figure~\ref{rhot}).}
\label{rt}
\end{center}
\end{figure}

\begin{figure}[t]
\begin{center}
\scalebox{0.75}{\includegraphics{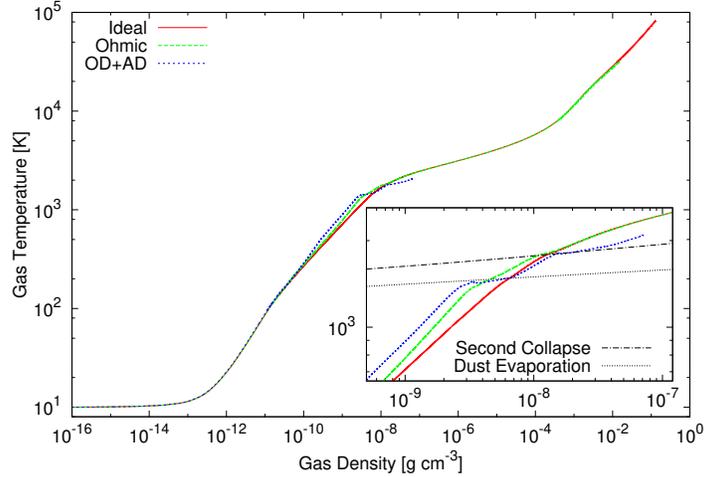}}
\caption{Evolution tracks of the gas elements at the center of the clouds in the $\rho$-$T$ plane. The small inset shows the region near the onset of the second collapse. The gray-dotted line indicates the temperature at which dust grains evaporate completely \citep{semenov}. The evaporation begins below this temperature and the opacity is already reduced. Also, the dash-dotted line roughly shows the temperature where the second collapse begins (i.e. $\gamma\ltsim 4/3$, see also \citet{tomida14}).}
\label{rhot}
\end{center}
\end{figure}

\begin{figure}[ht]
\begin{center}
\scalebox{0.75}{\includegraphics{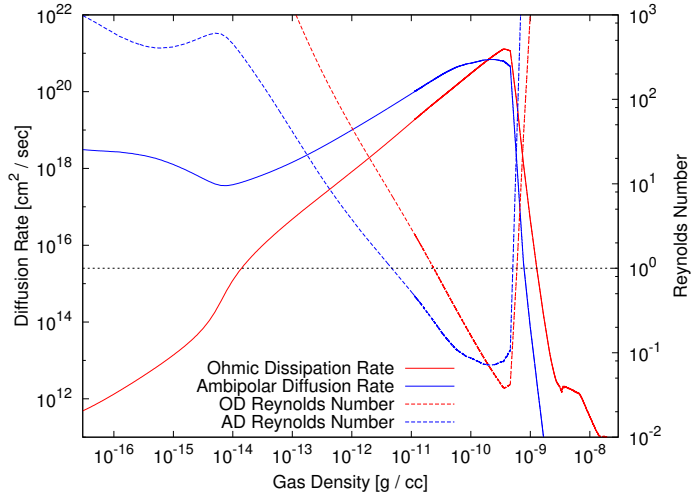}}
\caption{The rates and corresponding Reynolds numbers of Ohmic dissipation and ambipolar diffusion at the center of the cloud in Model {\it OA} as functions of the gas density. The magnetic Reynolds numbers are calculated assuming that typical velocity and length scale are the free-fall velocity and Jeans length; $R_{\rm O,A}\equiv v_{\rm ff}\lambda_{\rm J}/\eta_{O,A}$ (see Paper I, Appendix C).}
\label{rate}
\end{center}
\end{figure}

\begin{figure*}[t]
\begin{center}
\scalebox{0.81}{\includegraphics{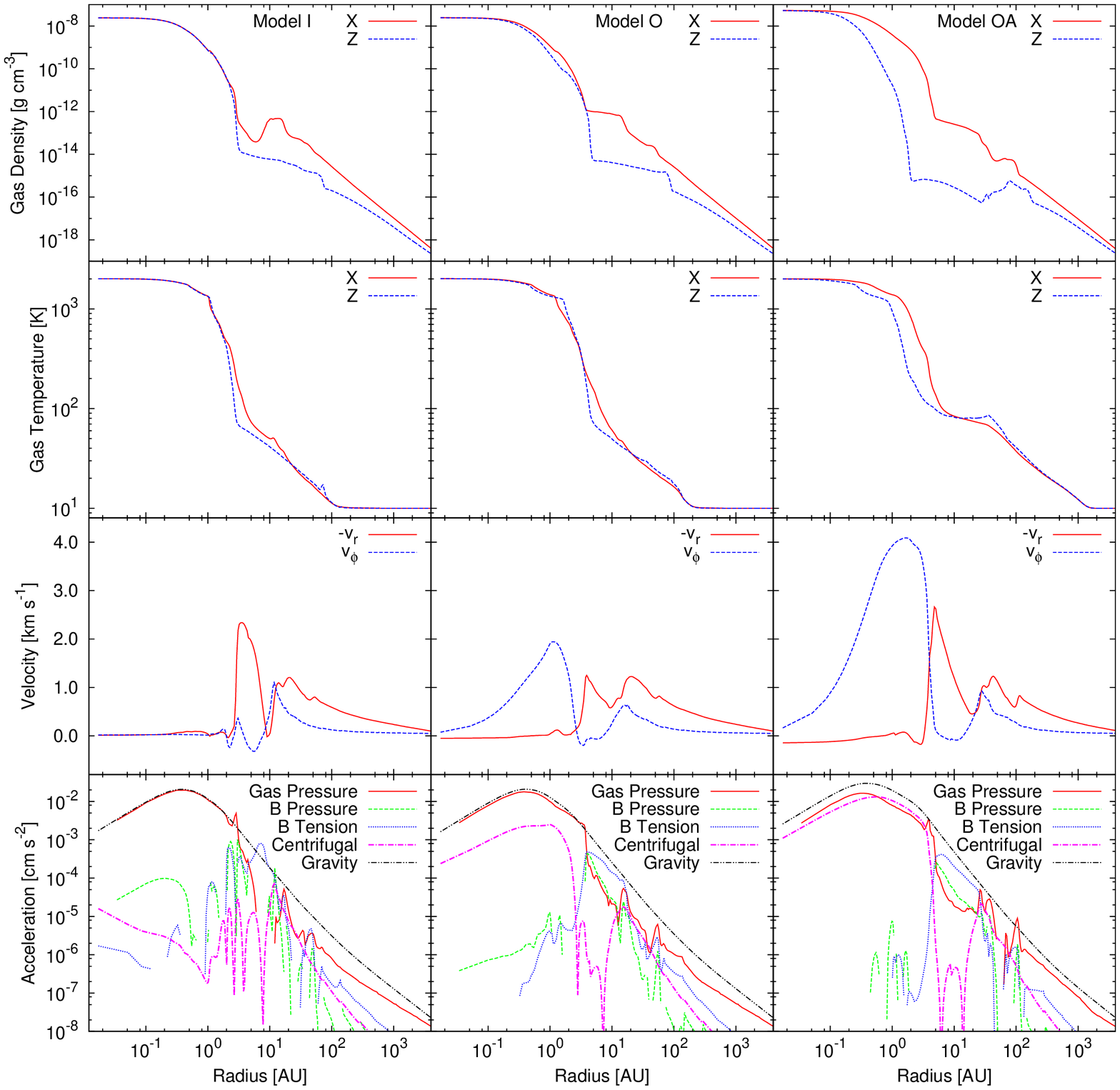}}
\caption{Radial profiles of the gas density, temperature along the $x$- (in the disk mid-plane, red) and $z$-axes (along the rotational axis, blue), the infall (red) and rotation (blue) velocities along the $x$-axis (in the mid-plane), and the radial accelerations by the various forces (The gas pressure gradient (red), the magnetic pressure gradient (green), the magnetic tension (blue), the centrifugal force (magenta), and the gravity (black). The radiation force is not plotted because it is negligibly small here) in the mid-plane at the end of the first core phase, from top to bottom. From left to right, the different columns are for Model {\it I}, {\it O} and {\it OA}, respectively.}
\label{prof}
\end{center}
\end{figure*}

\subsection{Overview of Evolution}
Here we summarize the evolutions of the three models. The time evolution of the density and temperature at the center of the clouds as well as the first core masses are shown in Figure~\ref{rt}. Figure~\ref{rhot} is the evolution tracks of the central gas element in the $\rho$-$T$ plane, and Figure~\ref{rate} shows the rates of Ohmic dissipation and ambipolar diffusion at the center of the cloud in Model {\it OA}. The profiles of the density, temperature, velocities and various forces at the end of the first core phase (we simply define it as $T_c=2,000\, {\rm K}$ although the effective adiabatic index gets below the critical value $\gamma=4/3$ slightly below this temperature) are shown in Figure~\ref{prof}. Model {\it I} is essentially the same as Model {\it IF} in Paper I, and Model {\it O} is similar to {\it RF} in Paper I.

 Initially the clouds collapse isothermally because radiation cooling is very effective, but the gas starts to evolve almost adiabatically when the central density exceeds about $\rho_c\sim 10^{-13} \, {\rm g \, cm^{-3}}$. Then the temperature quickly rises and quasi-hydrostatic objects, the first cores, are formed \citep{lrs69,mi00}. The formation of the first core is slightly earlier in Model {\it OA}, because ambipolar diffusion redistributes the magnetic flux before the gas becomes dense enough to form a first core, and it slightly weaken deceleration of the infall motion by magnetic fields. However, this is not significant because the dynamical collapse is much faster than the redistribution of the magnetic fields, and the difference is only 1\% compared to the time from the beginning of the collapse to the formation of the first core. It should be noted that this difference depends on the stability of the initial condition. If the initial condition is sufficiently unstable (super critical), ambipolar diffusion has only minor impact on the initial isothermal collapse \citep[see also][]{co14,hh14}.

Substantial differences arise during the first core phase. The first core evolves more slowly in the non-ideal MHD models, especially with ambipolar diffusion. The non-ideal MHD effects redistribute magnetic fluxes from the first cores and suppresses magnetic angular momentum transport. The additional rotational support changes the structures of the first cores and extends their lifetimes. Here we define the first core lifetime simply as the time between the epochs when the central temperature exceeds 100K and 2,000K. The lifetimes of the first cores are about 770 years in Model {\it I}, 1,050 years in {\it O}, but 2,750 years in {\it OA}. The first core in {\it OA} is qualitatively different and lives significantly longer than the others. While the accretion rates are essentially the same in all the models (initially $\dot{M}\sim4\times 10^{-5} \, {\rm M_\odot \, yr^{-1}}$ and gradually decrease later), the final masses are significantly different, about $0.03 \, {\rm M_\odot}$ in Model {\it I}, $0.04 \, {\rm M_\odot}$ in {\it O}, and $0.075 \, {\rm M_\odot}$ in {\it OA}. The first core in the ideal MHD model is essentially the same with spherical first cores without rotation, while the first core in Model {\it O} is slowly rotating but still looks almost spherical as rotational support is not dominant (Figure~\ref{prof}, see also Paper I). The first core in Model {\it OA} becomes disk-like and is clearly supported by rotation, although contribution of the gas pressure is still substantial. The non-ideal MHD effects also provide additional heating as in eq.(\ref{energy}), and the gas element follows a hotter (i.e. higher entropy) evolution track in Model {\it O}, and it is more prominent in {\it OA} (Figure~\ref{rhot}). These variations in the thermodynamic evolutions cannot be reproduced with the barotropic approximation. Because the hotter gas has higher pressure and can support more mass, this also contributes to the larger masses and longer lifetimes in the non-ideal MHD models.

When the temperature reaches about $1,400\, {\rm K}$, all the dust components evaporate and the opacities drop drastically (Figure~\ref{rhot}, see also Paper I). This effect is not obviously visible in the evolution track of Model {\it I} because it collapses very quickly. Model {\it O}, on the other hand, experiences considerable radiative cooling after dust evaporation and its evolution track merges to that of the ideal MHD model. In Model {\it OA}, the dust evaporation affects the thermal evolution more prominently because it evolves more slowly and therefore has more time to cool \citep[see also][]{tomida14}. It evolves along the dust evaporation temperature for a while\footnote{Although this behavior is interesting, this might be artificial because we use the tabulated dust opacities as functions of the density and temperature, assuming there is no hysteresis. Specifically, the evaporated dust grains immediately form again once the gas cools down below the evaporation temperature. In reality, it would take some time to reform dust grains and their structure and/or size distribution would be different from those before the evaporation. While this reformation of dust grains is highly uncertain, fortunately, it would not affect the evolution significantly. The gas evolves along the dust evaporation line if the opacities after reformation are higher. On the other hand, it evolves almost isothermally until the gas opacities and density become large enough if the opacities are lower, but the evolution is already close enough to isothermal.}, and the gas gets colder than those in Model {\it I} and {\it O}.

Dissociation of molecular hydrogen becomes significant when the temperature at the center of the clouds reaches about $T_c\sim 1,800\, {\rm K}$. Because of this strongly endothermic reaction, the gas pressure fails to maintain balance with the gravity and starts to collapse again. This second collapse is fairly violent in Model {\it I} and {\it O}. They dynamically collapses almost at the time scale of free-fall. After the second collapse completes, another hydrostatic objects, protostellar cores are formed. While the protostellar core in Model {\it I} is almost spherically symmetric, that in Model {\it O} is rotating and a disk is simultaneously formed (see also Paper I). Because the resistivity in this work is higher than that in Paper I, the disk is larger but its radius is still small at the end of the simulation, which is only $R \simeq 1\, {\rm AU}$. On the other hand, Model {\it OA} evolves again differently; because of the centrifugal force, the second collapse occurs rather gradually. Radiation cooling is still effective even in this phase and the gas evolution follows the colder track, deviating from the adiabatic evolution (Figure~\ref{rhot}). As mentioned above, we stop the calculations of Model {\it O} and {\it OA} earlier because fragmentation occurred.

The evolution of the diffusion rates and corresponding magnetic Reynolds numbers evaluated at the center of the cloud in Model {\it OA} is shown in Figure~\ref{rate}. The Ohmic dissipation rate in Model {\it O} behaves almost the same. The ambipolar diffusion rate is higher than the Ohmic dissipation rate in almost all the regions, and Ohmic dissipation becomes dominant over ambipolar diffusion only in the dense region where $\rho_c\gtsim 2 \times 10^{-10}\, {\rm g\, cm^{-3}}$. The Ohmic dissipation rate is very low in the low density region, and it becomes active only in the dense region within the first core ($2\times 10^{-11}\, {\rm g\, cm^{-3}} \ltsim \rho_c \ltsim 6\times 10^{-10}\, {\rm g\, cm^{-3}}$). The ambipolar diffusion rate is much higher than the Ohmic dissipation rate in the low density region, but it does not change the dynamics substantially until the first core is formed. Ambipolar diffusion works effectively in the broader range than the Ohmic dissipation ($5\times 10^{-12}\, {\rm g\, cm^{-3}} \ltsim \rho_c \ltsim 5\times 10^{-10}\, {\rm g\, cm^{-3}}$), removes magnetic flux and suppresses angular momentum transport more significantly. Both the Ohmic dissipation and ambipolar diffusion rates drop sharply when the gas temperature exceeds about $650 \, {\rm K}$ (or $\rho_c \sim 5\times 10^{-10}\, {\rm g\, cm^{-3}}$) because the gas and magnetic fields recouple by thermal ionization of potassium (K).

\subsection{Outflows}
\begin{figure*}[t]
\begin{center}
\scalebox{0.4}{\includegraphics{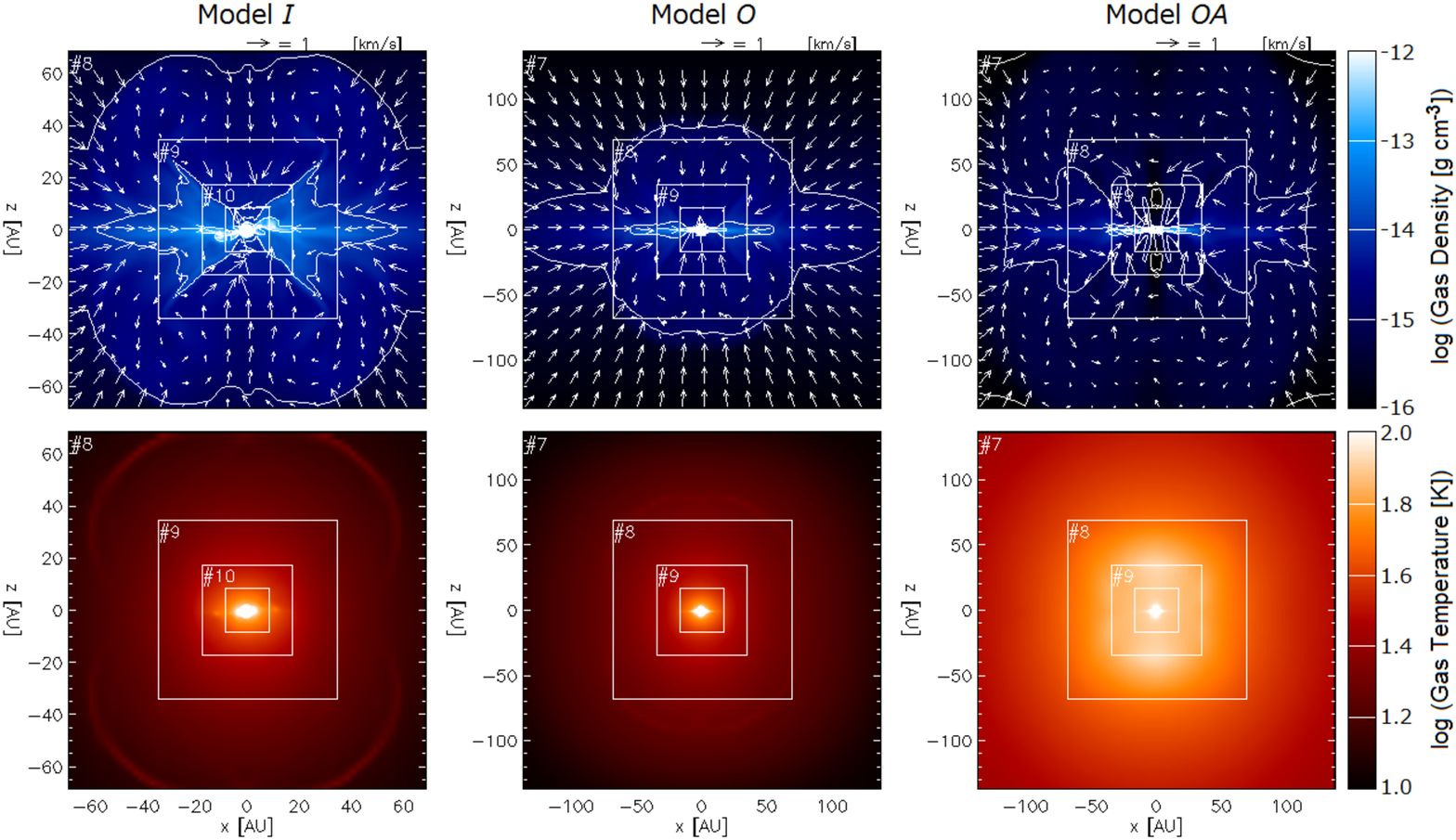}}
\caption{Vertical cross sections of the gas density (top) and temperature (bottom) in the outflow scales ($l=8$ or $\sim 140 {\rm AU}$ for Model {\it I}, and $l=7$ or $\sim 280 {\rm AU}$ for the rest) at the end of the first core phase ($T_c = 2,000\, {\rm K}$). Projected velocity vectors are overplotted with white arrows. Because the outflows are roughly axisymmetric in the large scale, only vertical cross sections are shown here.}
\label{outflow}
\end{center}
\end{figure*}

First, we compare the outflows driven around the first cores. Vertical cross sections of the density and temperature in this scale are presented in Figure~\ref{outflow}. These outflows are driven by the magneto-centrifugal force \citep{bp82} and remove angular momenta from the central objects efficiently \citep[e.g.][]{tmsk98,tmsk02,tomida13}. The driving regions are located around $R = 10 - 30\, {\rm AU}$, corresponding to the outer-most peaks of the rotational velocity $v_\phi$ in Figure~\ref{prof}. The gas density at these radii is still low and neither Ohmic dissipation nor ambipolar diffusion works strongly yet. The difference of the driving radii between the models is simply originated from the ages of the systems, because accreting angular momenta get larger as the systems evolve.

The sizes of the outflows are significantly different between the models. Before the end of the first core phase, the outflows extend vertically about 70 AU in Model {\it I}, 80 AU in {\it O}, and 170 AU in {\it OA}. However, the structures and velocities of the outflows are similar, about $1\, {\rm km\, s^{-1}}$ in all the cases. This is because the non-ideal MHD effects, both Ohmic dissipation and ambipolar diffusion, only work in the high density region within the first core, and do not affect the driving regions which are located outside the first cores as discussed above (Figure~\ref{prof}). Therefore, the larger sizes in the non-ideal MHD models are solely consequences of the longer first core lifetimes by rotational support.

While the outflow structures are quite similar, the temperature distribution in this scale is prominently warmer in Model {\it OA} than the other models. This is because the first core in {\it OA} is more massive and hotter, and heats up the surrounding material by radiation more strongly. Because of the longer lifetime and higher luminosity, the probability of finding a first core by observations would be higher than previously expected, although it should be still very rare because the lifetime is still much shorter than the timescale of the whole star formation process.

Another noticeable difference in the outflows is that a polar cavity with a very low density is formed in Model {\it OA}. The gas near the pole is removed by accretion and the centrifugal force, rather than expelled by the outflow itself. The opening angle of this cavity is still very narrow in the early phase, but it will expand as the gas in the outer envelope with higher angular momenta will accrete later. Although it is not very likely, this cavity might enable us to observe the central object more directly if it is in a face-on configuration.

\subsection{First Cores and Second Collapse}
\begin{figure}[t]
\begin{center}
\scalebox{0.8}{\includegraphics{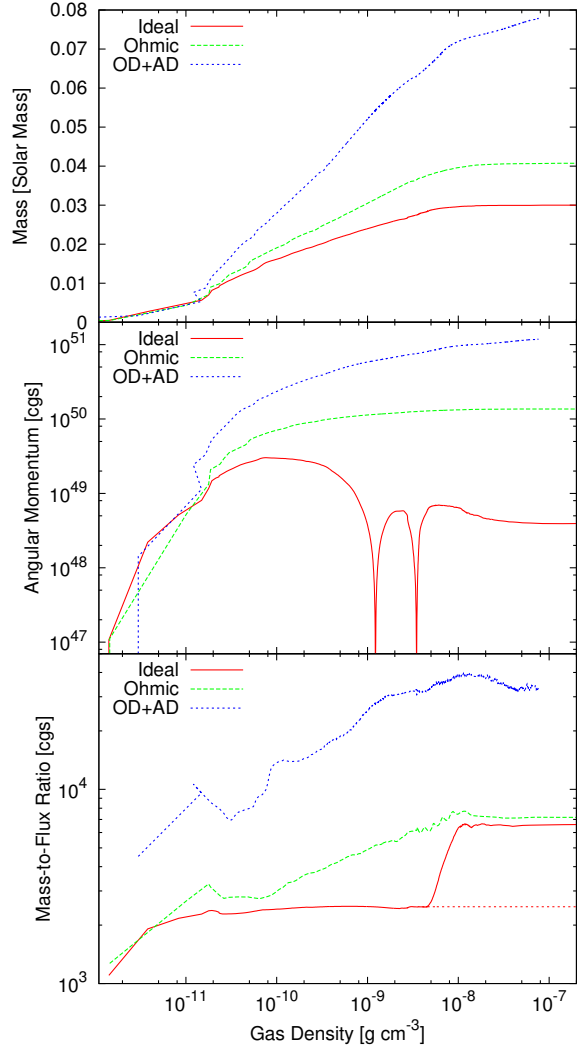}}
\caption{Evolution of the masses, angular momenta, and mass-to-flux ratios of the first cores from top to bottom as functions of the gas density at the center of the cloud. The red dotted line in the bottom panel is extrapolation of the mass-to-flux ratio before the magnetic interchange instability develops (see the text). Note that this mass-to-flux ratio is presented in the cgs unit and not normalized.}
\label{fcprop}
\end{center}
\end{figure}

As we can see in Figure~\ref{rate}, the non-ideal MHD effects are most active in the dense region within the first cores and significant magnetic flux loss occurs where the Reynolds number gets below unity. As a result, much more prominent differences between the models arise in the first core evolution and structures. Figure~\ref{fcprop} shows evolution of the masses, angular momenta and mass-to-flux ratios of the first cores. In order to produce these graphs, we define the first core by a simple criterion; where the gas density is higher than a critical density\footnote{This value is one order of magnitude higher than the critical density used in Paper I. We have changed the threshold so as to exclude the high density region in the pseudo-disk (a flattened disk-like structure but not supported by gas pressure or rotation) especially in Model {\it OA}. This is why the angular momentum of Model {\it I} shown in Figure~\ref{fcprop} behaves differently from that in Paper I, but in fact they are consistent if the same threshold is adopted.} $\rho_{\rm crit} = 10^{-12} \, {\rm g\, cm^{-3}}$. Also, the density and temperature cross sections of each model are shown in Figures~\ref{modeli}-\ref{modeloa}. Three epochs are shown in these Figures; just before thermal ionization of potassium ($T_c=600\, {\rm K}$, left), beginning of dust evaporation ($T_c=1,400\, {\rm K}$, middle) and the end of the first core phase, just after the onset of the second collapse ($T_c=2,000\, {\rm K}$, right). The magnetic field lines at the end of the first core phase are presented in Figure~\ref{bf}. The properties of the first cores and disks are summarized in Table~1. 

\subsubsection{Ideal MHD Model}
\begin{figure*}[p]
\begin{center}
\scalebox{0.435}{\includegraphics{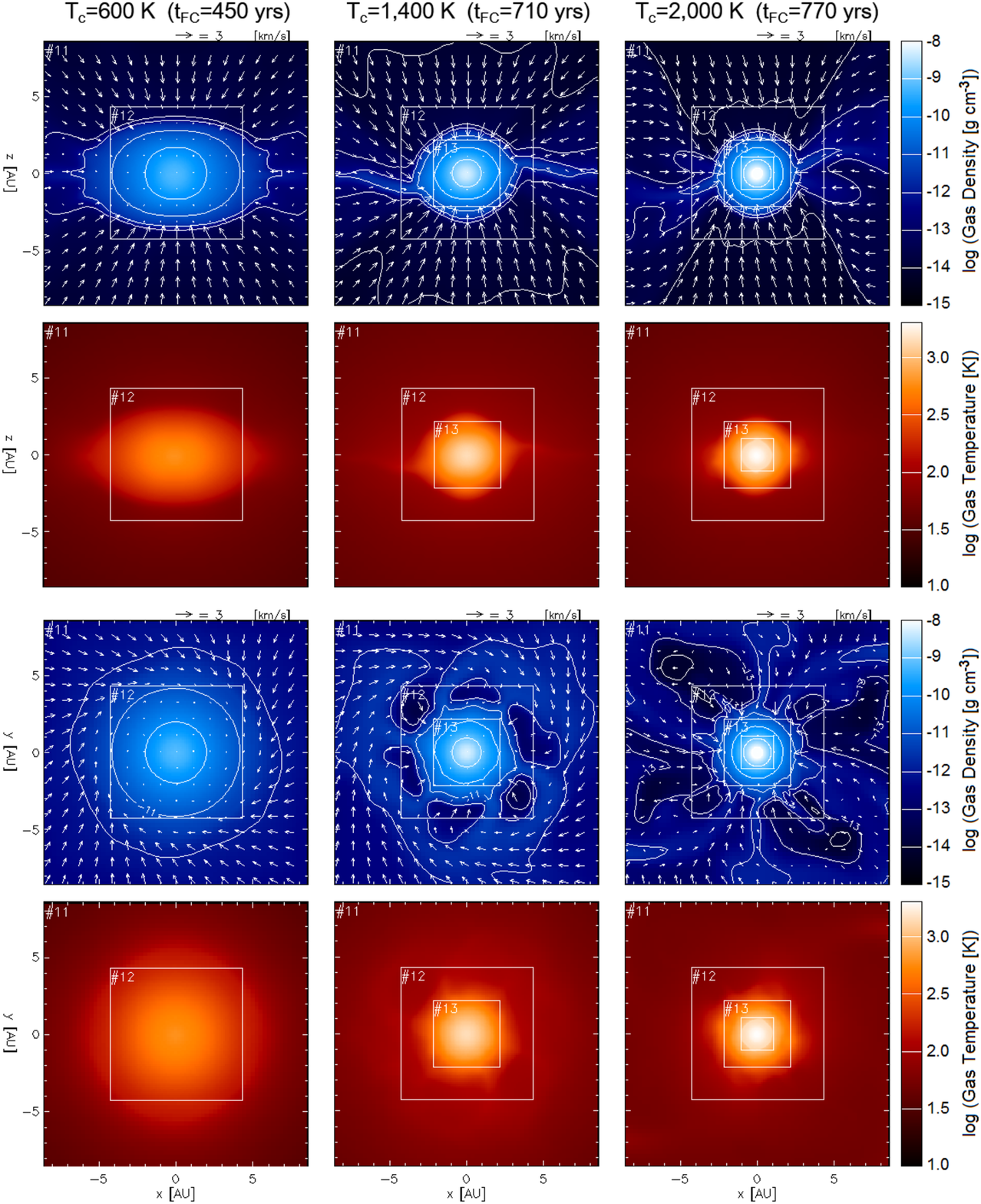}}
\caption{Vertical (top two rows) and horizontal (bottom two rows) cross sections of the gas density (the first and third rows) and temperature (the second and fourth rows) in the first core scale (l=11 or $\sim 18 {\rm AU}$) of Model {\it I}, when the temperature at the center of the cloud is 600K (left column), 1,400K (middle) and 2,000K (right). $t_{\rm FC}$ is the age of the first core after its formation (when $T_c=100\, {\rm K}$).}
\label{modeli}
\end{center}
\end{figure*}

The first core in Model {\it I} is essentially the same as that in Model {\it IF} of Paper I. Magnetic braking is so efficient that almost all the angular momentum in the first core is removed, and the first core is virtually spherically symmetric and not rotating (Figures~\ref{prof} and \ref{modeli}). The first core mass and radius at the end of the first core phase are about $0.03 M_\odot$ and 3 AU. The first core and outer pseudo-disk exhibit strong warping, which is driven by the magnetic interchange instability \citep{spr95,ss01,kras12}. The angular momentum in the first core decreases as it evolves. The two sharp ``wedges" are induced by strong perturbation from the magnetic interchange instability, indicating that the first core angular momentum is very small (essentially zero) and is easily dominated by the external warping. Because the thermal pressure and rotational support are smaller in this model than in the others, and the magnetic pressure is not significant, the first core mass is the smallest in this model.

To see the degree of magnetization, we look into the mass-to-flux ratio $M_{\rm FC} / \Phi_{\rm FC}$ of the first cores. The magnetic flux threading a first core is measured by integrating the vertical magnetic flux density over the mid-plane as follows:
\begin{equation}
\Phi_{\rm FC} \equiv \int_{\rm \rho>\rho_{\rm crit}} B_z dS = \sum_{\rho>\rho_{\rm crit}} B_z \Delta S,
\end{equation}
where $\Delta S = \Delta x \Delta y$ is the surface area of a discretized cell, and the summation is taken over the mid-plane. This formula captures the total magnetic flux correctly when the mid-plane is the plane of symmetry. In Model {\it I}, the symmetry breaks down by the magnetic interchange instability after the central density exceeds a few $\times 10^{-9}\, {\rm g \, cm^{-3}}$, while the other two models maintain the symmetry during the first core phase. Therefore, the mass-to-flux ratio in Model {\it I} after the sharp increase is not accurate, and it only gives the upper limit. Instead, it is more appropriate to assume this value is almost unchanged beyond this point (the red-dotted line in Figure~\ref{fcprop}) because the time after this is very short and the first core acquires only a little mass (see Figure~\ref{rt}). Taking this point into account, the first core magnetization remains almost constant during the first core phase in the ideal MHD model.

In the large scale, the magnetic fields look like the so-called hourglass shape. The field lines are tangled near and within the first core because of the magnetic interchange instability (Figure~\ref{bf}). Because the first core is not rotating, the field lines are not wound up. 

After the second collapse begins, the gas collapses dynamically and a protostellar core is formed very quickly in a few years. The protostellar core is also virtually spherical because almost no angular momentum exists in the protostellar core (at least shortly after its formation). The protostellar core properties are essentially the same as a spherical model without rotation and magnetic fields (Paper I).

\subsubsection{Non-Ideal MHD Model with Ohmic Dissipation Only}
\begin{figure*}[p]
\begin{center}
\scalebox{0.435}{\includegraphics{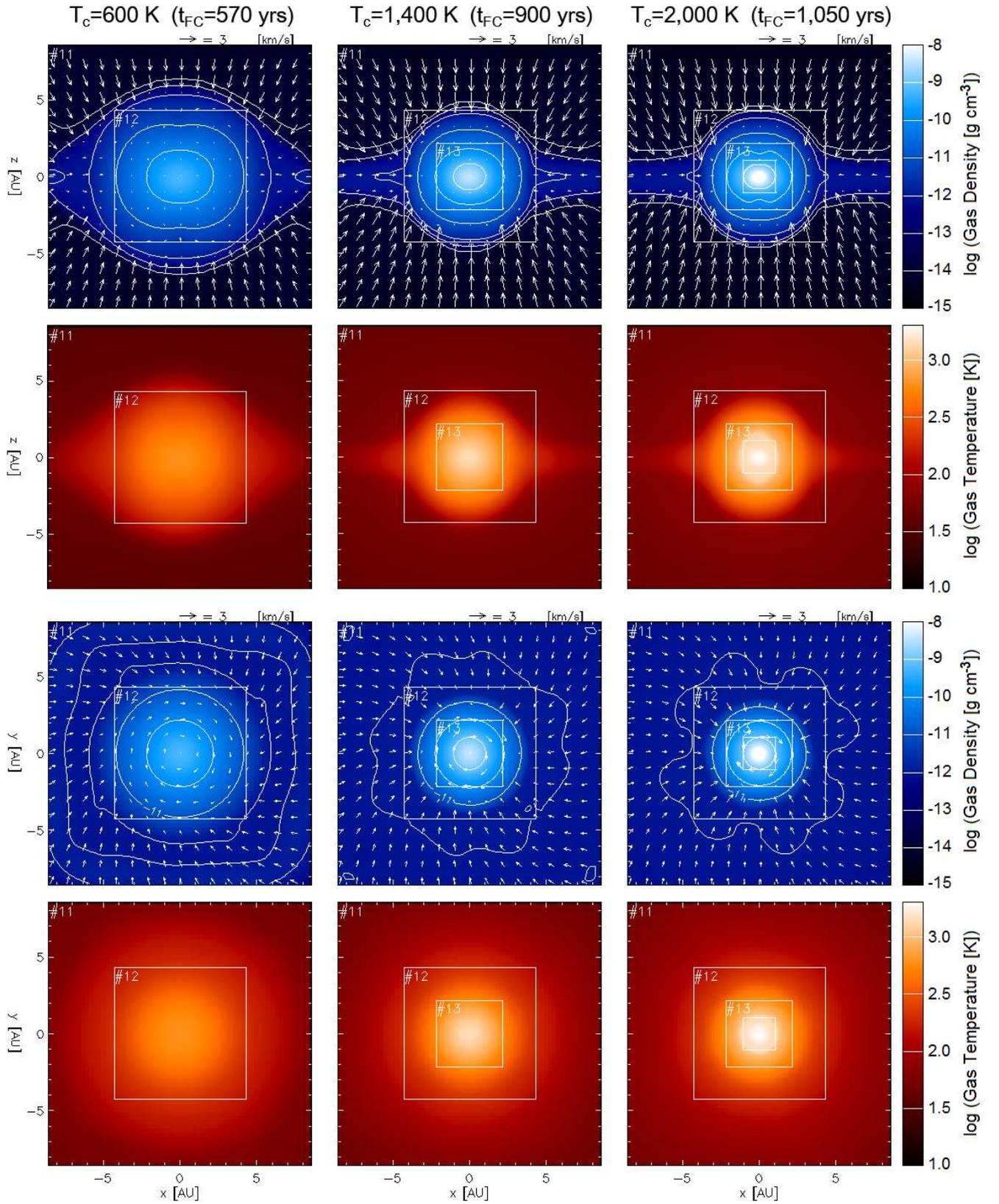}}
\caption{Same as Figure~\ref{modeli} but of Model {\it O}.}
\label{modelo}
\end{center}
\end{figure*}

Model {\it O} is very similar to Model {\it RF} in Paper I, although the Ohmic dissipation rate in this work is higher than the one used in Paper I. Distribution of dissipative regions are shown in Figure~\ref{reynolds} (the top row) using the magnetic Reynolds number $R_{\rm O}\equiv|\mathbf{v}|\lambda_{J}/\eta_{\rm O}$ where $\lambda_{\rm J}$ is the local Jeans length. Because Ohmic dissipation works only in the high density region within the first core, the first core properties (Figure~\ref{fcprop}) are identical to those in Model {\it I} when it is formed. As the first core evolves, the magnetic flux is redistributed outwardly by Ohmic dissipation. The gas and magnetic fields recouple in the central region after the thermal ionization of potassium begins, but the outer region remains resistive and the mass-to-flux ratio increases monotonically. At the end of the first core phase, the mass-to-flux ratio is about three times higher than in Model {\it I}. Angular momentum transport by magnetic fields is significantly suppressed, and the first core acquires a large angular momentum, which also monotonically increases. As a result, the first core has considerable rotation and evolves slightly more slowly than Model {\it I} (Figure~\ref{rt}). However, the rotation is not fast enough to support the first core and its structure remains almost spherical (Figures~\ref{prof} and \ref{modelo}). At the end of the first core phase, the mass and radius are about 0.04 $M_\odot$ and 5 AU, which are larger than the ideal MHD case because of additional support by rotation and heat produced by Ohmic dissipation (Figure~\ref{rhot}, see also Paper I). However, these differences in the first cores are rather minor. Thus, roughly speaking, Ohmic dissipation does not have a drastic impact on the first core structure and evolution. 

The magnetic fields in the outer region of the first core are straightened out by the strong Ohmic resistivity. On the other hand, the magnetic fields are tightly wound up and amplified by the shearing rotation in the hot recoupled region near the center of the first core (Figures~\ref{bf} and \ref{reynolds}). The magnetic pressure in this region is not strong enough to launch outflows during the first core phase. Later in the protostellar core phase, strong toroidal magnetic fields are produced by the faster rotation in the disk and drive fast jets by the magnetic pressure (see Paper I for the detailed discussion about the protostellar cores).

The second collapse occurs dynamically at the beginning as in Model {\it I} because the rotational support is not significant. However, since the gas spins up as it collapses, a rotationally supported disk forms around the protostellar core essentially in parallel. The disk size is very small, only about $1 \, {\rm AU}$ at the end of the simulation (when $T_c=30,000\, {\rm K}$). At the same time, fast bipolar jets are launched by the magnetic pressure of the toroidal magnetic fields amplified by the shearing rotation \citep[Paper I, see also][]{mim08}. This disk is larger than that in Paper I as a consequence of the higher Ohmic resistivity, but qualitatively the same. In this sense, Ohmic dissipation resolves the magnetic braking catastrophe and enables formation of a rotationally supported disk in the early phase of star formation. We stop the simulation because the Jeans condition cannot be satisfied beyond this point due to the limitation of the simulation code.

\subsubsection{Non-Ideal MHD Model with Ohmic Dissipation and Ambipolar Diffusion}
\begin{figure*}[p]
\begin{center}
\scalebox{0.435}{\includegraphics{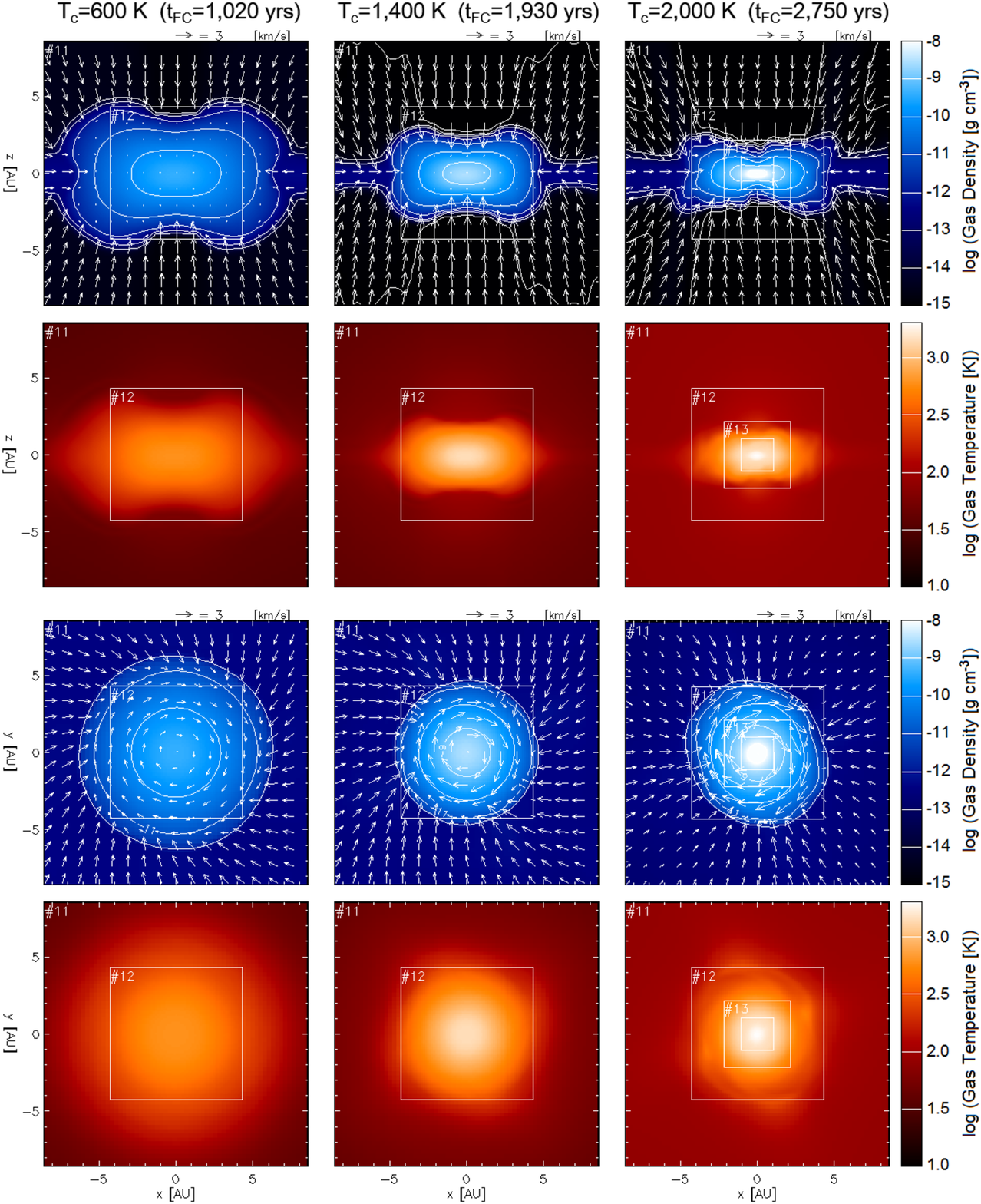}}
\caption{Same as Figure~\ref{modeli} but of Model {\it OA}.}
\label{modeloa}
\end{center}
\end{figure*}

\begin{figure*}[t]
\begin{center}
\scalebox{0.2}{\includegraphics{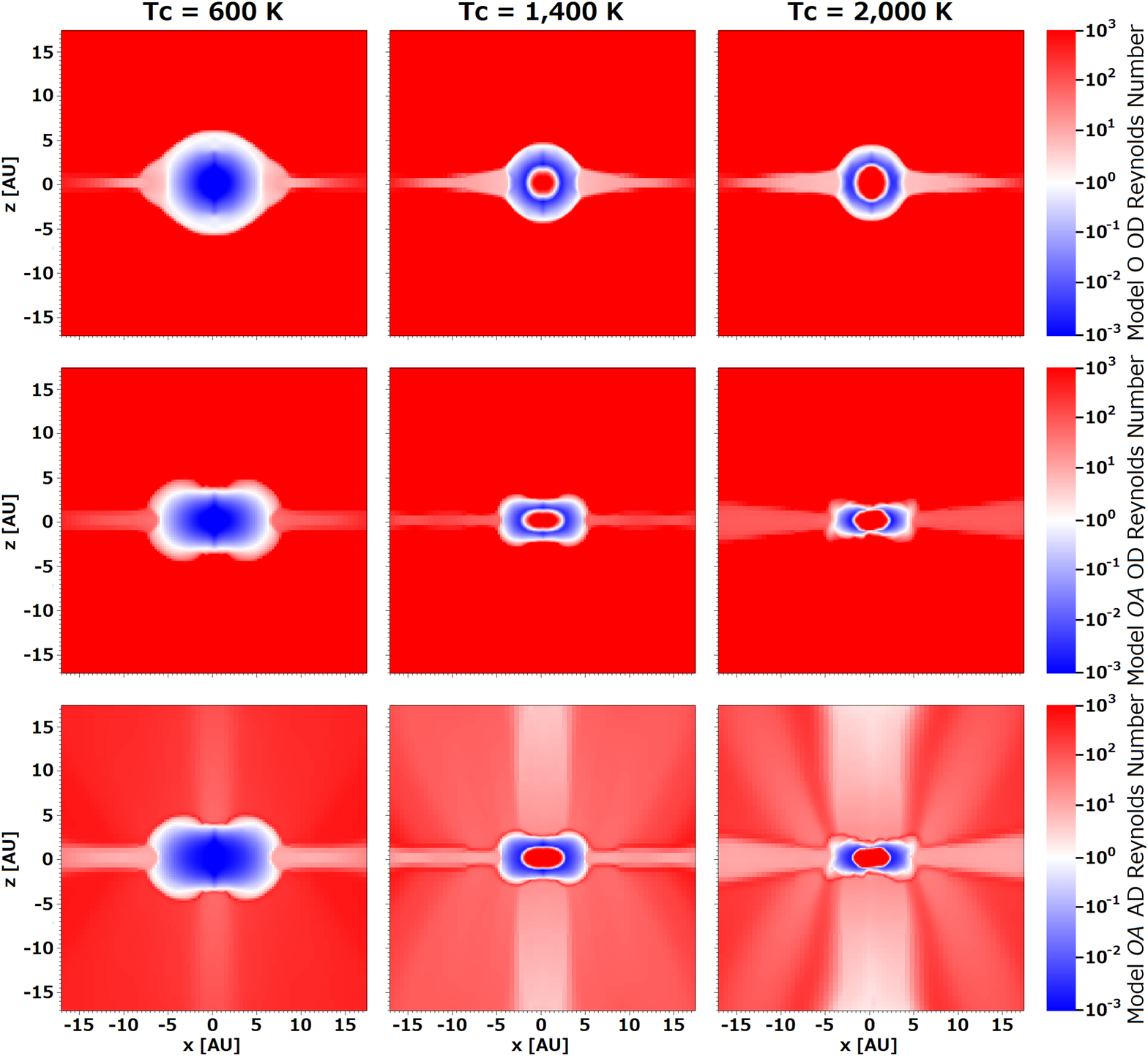}}
\caption{Vertical cross sections of the magnetic Reynolds numbers in the first cores and surrounding envelopes (l=10 or $\sim 35 {\rm AU}$), when the temperature at the center of the cloud is 600K (left column), 1,400K (middle) and 2,000K (right). We use the gas velocity and local Jeans length to evaluate the Reynolds numbers. The top row shows the Reynolds number calculated from the Ohmic resistivity in Model {\it O}. The bottom two rows are Model {\it OA}, while the middle row indicates the Ohmic Reynolds number and the bottom row is calculated using the ambipolar diffusion rate. The non-ideal MHD effects work effectively in the blue regions while the red regions are close to ideal.}
\label{reynolds}
\end{center}
\end{figure*}

The first core in Model {\it OA} with both Ohmic dissipation and ambipolar diffusion is qualitatively different from the other two cases. Ambipolar diffusion is more effective than Ohmic dissipation especially in the low density gas (Figure~\ref{rate}), and suppresses angular momentum transport more significantly. The first core is already less magnetized from the beginning, and it loses the magnetic flux continuously during the first core phase (Figure~\ref{fcprop}). Figure~\ref{reynolds} indicates that the non-ideal MHD effects work strongly in almost the whole first core, especially in the early phase \citep[see also][]{susa15}. At the end of the first core phase, the mass-to-flux ratio is higher than the ideal MHD model by more than one order of magnitude.

Angular momentum transport by magnetic fields is strongly suppressed in such a diffusive region, and substantial angular momentum remains within the first core. As a result, the first core becomes a disk supported by the centrifugal force, although the gas pressure is still substantially contributing (Figure~\ref{prof}). The rotational profile is not Keplerian yet (i.e. the rotation profile is not $v_\phi \propto R^{-1/2}$) because of this gas pressure support and the self-gravity. The disk radius is initially about 6 AU when it is formed, gradually shrinks to $\sim 5\, {\rm AU}$, and then remains almost constant, because magnetic braking is still working outside the resistive region and regulates angular momentum accretion. Although the disk remains small until the end of the simulation, it accumulates large angular momentum as the gas accretes from the envelope. Eventually it acquires almost one order of magnitude larger angular momentum than that in Model {\it O}, while the mass ($0.075M_\odot$ at the end of the first core) is larger only by a factor of two. This fast rotation is sufficient to the trigger the gravito-rotational instability \citep{toomre,bate98,saigo06,saigo08,tm11}, and the disk becomes non-axisymmetric in the late phase (Figure~\ref{modeloa}). Angular momentum transport by gravitational torque via non-axisymmetric structures is usually less effective than magnetic braking, but it becomes important where the magnetic fields are weakened by the non-ideal MHD effects. It is naturally expected that this disk will grow later and the rotational support will be more significant as the gas with higher angular momenta will accrete from the outer region.

The magnetic flux expelled by the non-ideal MHD effects piles up outside the diffusive region ($R\sim 5 \,{\rm AU}$), i.e. the first core. In this situation, the so-called magnetic wall can be formed, where the magnetic pressure strongly decelerates the infalling gas and possibly form a shock \citep[e.g.][]{lm96,tm05a,tm05b,tm07,kunz10}. However, such a structure is not prominent in this model, although the accretion velocity is indeed slower than the free-fall velocity due to the magnetic fields (Figure~\ref{prof}). The outermost deceleration (at $R\sim 100\, {\rm AU}$) is formed by interacting with the bipolar outflow (see Figure~\ref{outflow}), the second one (at $R\sim 30\, {\rm AU}$) is a centrifugal barrier, and the innermost one (at $R\sim 5\, {\rm AU}$) is the first core shock. This is likely because our rates of the non-ideal MHD effects are not high enough, or because the simulation duration is not long enough to accumulate sufficient magnetic flux.

As discussed before, this model suggests that first cores are more readily observed because of the higher luminosity and longer lifetime. Another interesting observational aspect is its variability. While mass and angular momentum transport by magnetic fields are almost steady and smooth, the gravitational torque via non-axisymmetric structures is unsteady and episodic \citep[see also][]{vb06,vb10,stm11}. When the disk becomes gravitationally unstable by mass accretion or radiation cooling, non-axisymmetric structures like spiral arms are formed and transport mass and angular momentum, then the disk is stabilized again unless it is too unstable and fragments. This stochastic accretion causes the first core flickering. Its time scale is corresponding to the orbital time scales, ranging from a few years to a few decades. We will discuss its observational properties in a separate paper in the near future.

\begin{figure}[t]
\begin{center}
\scalebox{0.32}{\includegraphics{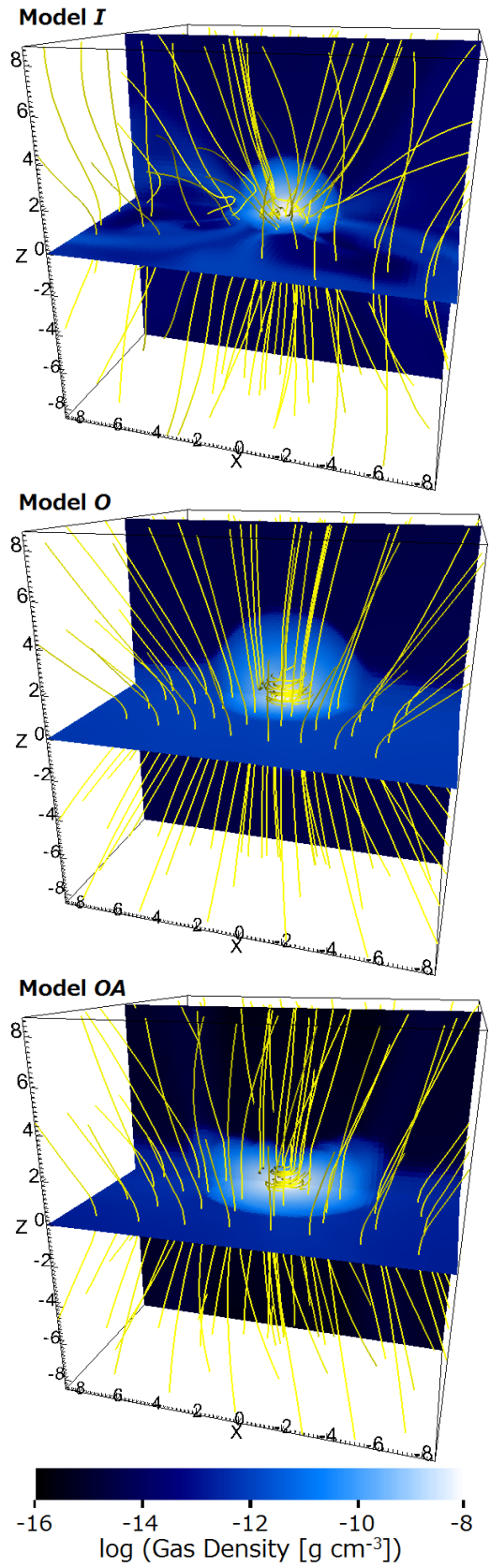}}
\caption{Magnetic field lines (yellow lines) threading the first cores and surrounding pseudo-disks (l=11 or $\sim 18\, {\rm AU}$) at the end of the first core phase. Vertical and horizontal density cross sections are also presented. Note that the density of the field lines does not reflect the strength of the magnetic fields.}
\label{bf}
\end{center}
\end{figure}

Similar to the case with Ohmic dissipation only, the magnetic field lines in the outer region of the first core are strongly straightened out, running almost vertically, by strong Ohmic dissipation and ambipolar diffusion (Figures~\ref{bf} and \ref{reynolds}). The field lines are wound up tightly near the center of the first core where the magnetic fields are coupled with the gas, but again the magnetic pressure is not dominant during the first core phase and the fields behave passively in this phase.

Although we stop the simulation soon after the second collapse begins (when $T_c \sim 2,100 \, {\rm K}$), we can see that the evolution during the second collapse is already different from the other two models. Because it is supported by the centrifugal force, the gas does not collapse dynamically even when it loses the gas pressure support. It rather collapses gradually with a time scale much longer than the free fall. Unfortunately, we have to stop the simulation at this point because the Jeans condition is violated and fragmentation occurred. We cannot tell but only can speculate how the system will evolve later, but it is likely that a compact binary or multiple system is formed \citep{mtmi08}.

To summarize, ambipolar diffusion is so effective and drastically changes the first core structure. It removes magnetic flux from the first core, suppresses angular momentum transport, and enables early formation of rotationally supported disks, even before formation of the protostar. The disk is massive enough to become gravitationally unstable and becomes non-axisymmetric. On the other hand, the disk size remains small in the early phase, contrary to purely hydrodynamic models without magnetic fields which yields very large disks supported by rotation even in the first core phase \citep{bate98,bate10,saigo08,tm11,tomida14}. Thus, these non-ideal MHD effects can solve the magnetic braking catastrophe and maintain consistency with observations of young stellar objects \citep[e.g.][]{maury}.

\section{Discussion}
\subsection{Comparison with the Precedent Non-Ideal MHD Simulations}
Our results seem to be inconsistent with the earlier non-ideal MHD simulations by \citet{li11} \citep[see also][]{ml09}, in which no rotationally supported disk was formed even with ambipolar diffusion. Because there are many differences in the simulation setups including the initial conditions, comparison is not trivial but at least the diffusion rates are similar. We suspect that the most important difference between their simulations and ours is the boundary condition at the center of the cloud. They have a relatively large hole with a radius of 6.7 AU, while we do not have any boundary there but resolve the small-scale structure directly using the nested-grid technique. In fact, because the disks formed in our simulations are small, it is not inconsistent with their simulations, at least in the early phase of protostellar collapse.

Our calculations are expensive and we cannot follow the long-term evolution, but this high resolution is required to resolve the small first core disk. We believe it is crucial to resolve the first core because the non-ideal MHD effects work effectively and a rotationally supported disk is formed there. Otherwise, these non-ideal MHD effects have only minor influence on the large scale evolution. Indeed, \citet{machida14} demonstrated that the sink particle or inner hole has a significant impact on the results. We propose that the effects of these numerical tricks can be estimated by measuring the angular momentum accreted into the inner boundary or sink particles.

\subsection{Perspectives of Future Observations}
As discussed in the introduction, recent interferometric observations of Class-0 objects, the youngest protostars including L1527 IRS, suggest that circumstellar disks in the early phase of star formation should be small, i.e. $R\ltsim 100\, {\rm AU}$, while they can be formed in the early phase \citep{maury,ohashi}. These observational results do not favor previous hydrodynamic simulations of protostellar collapse without magnetic fields which tend to produce large rotationally supported disks in the early phase \citep[e.g.][]{bate98,bate10,saigo08,tm11}. In contrast, our model with both ambipolar diffusion and Ohmic dissipation is consistent with this picture. In order to test the scenario, more observations of very young stellar objects with higher resolutions that can sufficiently resolve the circumstellar disks are required. In particular, if we could directly observe a first core or at least its remnant shortly after the second collapse, it would be a critical test for our models. If we could detect a compact rotating disk in a collapsing cloud but associated with no significant stellar signature (e.g. near infrared emission or high velocity jets), it would be crucial evidence indicating that a rotationally supported disk is already formed in the first core phase.

From numerical simulations \citep[e.g.][]{by95,tomida10b,tt11,st11,com12}, a first core is expected to (1) have a spectral energy distribution like a low-temperature gray body without a hot infrared stellar emission, (2) be a compact, condensed core which may exhibit a signature of infall motion, and (3) be associated with slow compact molecular outflows, but no fast jets. Although first cores should be very rare due to the short lifetime that only one out of a few hundreds of molecular cloud cores are expected, some candidates are already reported \citep{chen10,en10,chen12,pezzuto12,hirano14,tokuda14}. Among them, L1451-mm \citep{pineda11} is one of the best examples which satisfies all the above criteria. We expect that direct observation of a first core or its remnant with high resolution interferometers, especially ALMA, will be achieved in the near future.

\subsection{Non-Ideal MHD Effects}
While the non-ideal MHD effects can avert the magnetic braking catastrophe, the magnetic flux problem may remain a serious problem. As shown in Figure~\ref{fcprop}, the non-ideal MHD effects increase the mass-to-flux ratio of the first core but only by one order of magnitude. This is insufficient to explain the difference of magnetization between molecular cloud cores and protostars. This is partly because we do not consider shielding of cosmic rays \citep[see also][]{dapp12,padovani14}, while the ionization rate in the dense region would largely vary depending on the abundance of radioactive nuclei as discussed in \ref{srate}. Possibly, higher non-ideal MHD rates or other mechanisms to accelerate diffusion (e.g. turbulence) are required. If we use higher rates, the disk size will be also larger. Or, we speculate that further redistribution of magnetic fields from the formed protostar will take place in the later phase \citep[e.g.][]{bra12}.

The non-ideal MHD effects play an important role in numerical simulations; they actually help convergence. In the ideal MHD approximation, dissipation of magnetic fields is totally determined by resolution. In protostellar collapse simulations, the mid-plane is a current sheet where the magnetic fields above and below are anti-parallel. In ideal MHD, the current sheet can be extremely thin and convergence is not well defined, at least difficult to achieve. For example, \citet{joos12,joos13} reported poor convergence of their ideal MHD simulations with mis-aligned magnetic fields or turbulence; more than 20 cells per the local Jeans length are required for the mis-aligned cases while 15 cells seem to be sufficient for the aligned cases. With the non-ideal MHD effects, on the other hand, the current sheet has a finite thickness, and therefore convergence is now well defined and can be achieved with a finite resolution. However, it still requires high resolution to resolve the dissipation scales sufficiently because the dissipation scale is very small due to the small diffusivity in the low density region.

Although our simulations include as many physical processes as possible, a possibly important effect missing in this work is the Hall effect. The Hall effect is another non-ideal MHD effect which can spin up the gas and help disk formation \citep{kra11,li11,bw12}. Because the rate of the Hall effect is as high as the ambipolar diffusion \citep{wn99,nkn02,li11}, it is likely important in star and disk formation processes. However, it is not easy to implement in high-resolution 3D simulations, especially when we resolve the first core directly where the Hall effect works most actively. This is because it requires very small time steps proportional to $\Delta x^2/\eta_{\rm H}$ where $\eta_{\rm H}$ is the rate of the Hall effect. The STS acceleration is not effective for the Hall effect because of its hyperbolic nature. Moreover, unlike Ohmic dissipation and ambipolar diffusion, the Hall effect changes the characteristic speeds of the MHD waves (e.g. whistler waves). Therefore, to capture this effect accurately, it is inadequate in principle to use the operator-splitting and sub-cycling techniques for the Hall effect, like we did for Ohmic dissipation and ambipolar diffusion. Instead, we have to solve the MHD part and the non-ideal MHD part using the same, very small time steps required by the Hall effect, which will be extremely expensive.

Finally, it should be noted that there are significant uncertainties in the rates of the non-ideal MHD effects because of large uncertainties in the underlying dust properties as discussed in \ref{srate}. The uncertainties also exist in the dust opacities used in the radiation transfer part. Because these uncertainties have direct impacts on formation and evolution of first cores, the results, especially the disk size, might be easily changed by an order of magnitude. Therefore, this work should be considered as a case study based on simple assumptions on the dust properties. A broad parameter survey is required to clarify the effects of the dust properties such as the structure and size distribution. It will be important and interesting to solve evolution of dust grains, including both growth and destruction, consistently in parallel with RMHD simulations. There are many topics on dust left to be explored for the future works.

\section{Conclusions}
In order to study the effects of Ohmic dissipation and ambipolar diffusion on protostellar collapse and disk formation, we performed 3D nested-grid RMHD simulations. We simulated only the early phase of star formation before formation of protostellar cores, but we found that significant differences already arose at this point. We revealed that the non-ideal MHD effects qualitatively change the structure and evolution of the first cores. Especially, ambipolar diffusion expands the diffusive region and strongly suppresses angular momentum transport, resulting in early formation of rotationally supported disks. The conclusions are summarized as follows:
\begin{enumerate}
\item Ohmic dissipation works in the high density region and suppresses angular momentum transport. As a result, a very small rotationally supported disk is formed after the second collapse, essentially in parallel with formation of the protostellar core.
\item Ambipolar diffusion extends the diffusive region to the lower density gas, and suppresses angular momentum transport more effectively. The first core is already supported by rotation significantly from the beginning, although the thermal pressure is still important.
\item The formed disks remain small in the early phase of star formation and the rotation profile is not Keplerian yet, but they will grow later as gas accretion with larger angular momenta continues.
\item The first core disk with ambipolar diffusion acquires sufficiently large mass and angular momentum to become gravitationally unstable and form non-axisymmetric structures. As a result, the accretion becomes naturally episodic. Although this is speculative at this point, it is likely that disk fragmentation occurs and a binary/multiple system will be formed later.
\item Because of additional support by rotation, the first core is more massive, longer-lived, and more luminous in the model with ambipolar diffusion. This increases the probability of first core detections, although it is still rare. Also, the first core can exhibit variability by the gravitational instability with the short time scale on the order of years.
\end{enumerate}

These results can reconcile the observational results and theoretical studies. In previous hydrodynamic simulations of protostellar collapse without magnetic fields, large rotationally supported disks are formed in the early phase of star formation, but those results are not favored by the observations of very young stellar objects. In ideal MHD simulations, on the other hand, the magnetic braking is so efficient that a circumstellar disk is difficult to form, which is referred as the magnetic braking catastrophe. Our results suggest that the non-ideal MHD effects can resolve the magnetic braking catastrophe in the early phase of star formation even before a protostellar core is formed, and enable formation of a small disk in the early phase. It is expected that the disk will grow later as larger angular momenta accrete.

In previous studies of disk evolution and planet formation, simplified models of isolated, already-established protoplanetary disks like the minimum mass solar nebula \citep{mmsn} have been often used as the initial and boundary conditions. However, our results suggest that a seed of a circum``stellar" disk can be formed in the early phase even before a protostar is formed, it co-evolves with the central protostar, and disk fragmentation can happen while the disk and star are still growing via accretion. This means that we need to consider more realistic situations in order to understand disk evolution and planet formation. Ultimately, we need a consistent scenario of star, disk and planet formation -- or stellar system formation.

\acknowledgments
We are grateful to Professor Eve C. Ostriker, Professor James M. Stone, Professor Tomoaki Matsumoto, Dr. Matthew Kunz, Dr. Yasunori Hori and Dr. Yusuke Tsukamoto for fruitful discussions and encouraging comments. We also thank the anonymous referee for the useful comments to improve this paper. This work is partly supported by the Ministry of Education, Culture, Sports, Science and Technology (MEXT), Grants-in-Aid for Scientific Research 23103005, 25887023, 26400224 (SO), 25400232 and 26103707 (MNM). KT is supported by Japan Society for the Promotion of Science (JSPS) Research Fellowship for Young Scientists ($26\cdot6947$). This research used computational resources of the High Performance Computing Infrastructure (HPCI) system provided by the Cybermedia Center at Osaka University and the Cyberscience Center at Tohoku University through the HPCI System Research Project (Project ID: hp140065).

\appendix
\section{A. Numerical Methods for Ohmic Dissipation and Ambipolar Diffusion} \label{nm}
\subsection{A.1. Conservative Form}
In this appendix, we describe how to update the induction and energy equations related to Ohmic dissipation and ambipolar diffusion. We adopt the operator-splitting approach and solve these effects separately from the other physical processes. The equations solved in this part are as follows:
\begin{eqnarray}
\left.\frac{\partial\mathbf{B}}{\partial t}\right|_{\rm diss}+\nabla\times\left(\eta_{\rm O}\mathbf{J}+\frac{\eta_{\rm A}}{|\mathbf{B}|^2}\mathbf{B}\times\mathbf{F}\right)&=&0,\label{diss}\\
\left.\frac{\partial e}{\partial t}\right|_{\rm diss}+\nabla\cdot\left[\eta_{\rm O}\mathbf{F}+\frac{\eta_{\rm A}}{|\mathbf{B}|^2}(\mathbf{B}\times\mathbf{F})\times\mathbf{B}\right]&=&0,\label{disse}\\
\mathbf{J}\equiv \nabla\times \mathbf{B},\hspace{2em} \mathbf{F}\equiv \mathbf{J}\times \mathbf{B}.\nonumber
\end{eqnarray} 

We can rewrite eq.~(\ref{diss}) into a conservative form:
\begin{eqnarray}
\left.\frac{\partial \mathbf{B}}{\partial t}\right|_{\rm diss}+\nabla\cdot\mathbb{T}^{\rm diss}&=&0,
\end{eqnarray}
where $\mathbb{T}^{\rm diss}=\mathbb{T}^{\rm O}+\mathbb{T}^{\rm A}$, $\mathbb{T}^{\rm O}$ and $\mathbb{T}^{\rm A}$ are the flux tensors of Ohmic dissipation and ambipolar diffusion, respectively. The fluxes of $B_b$ component in $a$-direction are:
\begin{eqnarray}
T_{ab}^{\rm O}&=&-\eta_{\rm O}(\partial_a B_b-\partial_b B_a),\\
T_{ab}^{\rm A}&=&-\frac{\eta_{\rm A}}{|\mathbf{B}|^2}(B_a F_b-B_b F_a).
\end{eqnarray}
Note that the dot product ($\cdot$) of a vector and a tensor is defined as summation over the first suffix of the tensor; $\mathbf{v}\cdot\mathbb{T}\equiv v_a T_{ab}$ using the Einstein summation convention, which yields a vector. With these fluxes, considering that these flux tensors are anti-symmetric and their diagonal components are zero ($T_{aa}=0$), eq.~(\ref{disse}) is also rewritten like:
\begin{eqnarray}
\left.\frac{\partial e}{\partial t}\right|_{\rm diss}+\nabla\cdot(\mathbb{T}^{\rm diss}\cdot\mathbf{B})&=&0.
\end{eqnarray} 
Again note that the dot product ($\cdot$) of a tensor and a vector is summation over the second suffix of the tensor; $\mathbb{T}\cdot\mathbf{v}\equiv T_{ab}v_b $.

\subsection{A.2. Correction Terms for ${\rm div} \mathbf{B}$}
Magnetic fields are defined at the cell centers in our scheme. In order to satisfy the solenoidal constraint (eq.\ref{sole}), we use the mixed cleaning scheme \citep{dedner} in the MHD part. By introducing additional correction terms for the non-ideal MHD part, we can improve the nature of the equations and reduce the divergence error. For Ohmic dissipation, this is achieved using the method proposed by \citet{grv08}, adding correction terms proportional to ${\rm div} \mathbf{B}$ in the diagonal components of the Ohmic dissipation fluxes:
\begin{eqnarray}
T_{aa}^{\rm O}&=&-\eta_{\rm O}\nabla\cdot\mathbf{B}.
\end{eqnarray}
For the ambipolar diffusion part, this is done by introducing additional source terms \citep{dp08}:
\begin{eqnarray}
\left.\frac{\partial \mathbf{B}}{\partial t}\right|_{\rm corr}&=&-\frac{\eta_{\rm A}}{|\mathbf{B}|^2}\mathbf{F}\nabla\cdot \mathbf{B}\label{srcb},\\
\left.\frac{\partial e}{\partial t}\right|_{\rm corr}&=&-\frac{\eta_{\rm A}}{|\mathbf{B}|^2}(\mathbf{B}\cdot\mathbf{F})\nabla\cdot \mathbf{B}\label{srce}.
\end{eqnarray}

\subsection{A.3. Discretization}
Discretization of eqs.(\ref{diss}) and (\ref{disse}) is straightforward:
\begin{eqnarray}
\left.\frac{\partial B_a^{i,j,k}}{\partial t}\right|_{\rm diss}&+&\frac{T_{xa}^{i+1/2,j,k}-T_{xa}^{i-1/2,j,k}}{\Delta x}\nonumber\\&+&\frac{T_{ya}^{i,j+1/2,k}-T_{ya}^{i,j-1/2,k}}{\Delta y}\nonumber\\&+&\frac{T_{za}^{i,j,k+1/2}-T_{za}^{i,j,k-1/2}}{\Delta z}=0,\\
\left.\frac{\partial e^{i,j,k}}{\partial t}\right|_{\rm diss}&+&\frac{T_{xa}^{i+1/2,j,k}B_a^{i+1/2,j,k}-T_{xa}^{i-1/2,j,k}B_a^{i-1/2,j,k}}{\Delta x}\nonumber\\&+&\frac{T_{ya}^{i,j+1/2,k}B_a^{i,j+1/2,k}-T_{ya}^{i,j-1/2,k}B_a^{i,j-1/2,k}}{\Delta y}\nonumber\\
&+&\frac{T_{za}^{i,j,k+1/2}B_a^{i,j,k+1/2}-T_{za}^{i,j,k-1/2}B_a^{i,j,k-1/2}}{\Delta z}=0\label{dde},
\end{eqnarray}
where $i,j$ and $k$ are the spatial indices of the cell centers in the $x,y$ and $z$ direction, and $i\pm1/2$ means the location of the cell surface. The Einstein notation is used in eq.(\ref{dde}); $T_{xa}B_a=\sum_{a=x,y,z} T_{xa}B_a$, etc.. The fluxes at the cell surface $(i+1/2,j,k)$ are calculated using the following relations and their permutations:
\begin{eqnarray}
(\partial_x B_y)^{i+1/2,j,k}&=&\frac{B_y^{i+1,j,k}-B_y^{i,j,k}}{\Delta x},\\
(\partial_y B_x)^{i+1/2,j,k}&=&\frac{B_x^{i+1,j+1,k}+B_x^{i,j+1,k}-B_x^{i+1,j-1,k}-B_x^{i,j-1,k}}{4\Delta y},\\
\mathbf{B}^{i+1/2,j,k}&=&\frac{\mathbf{B}^{i+1,j,k}+\mathbf{B}^{i,j,k}}{2},\\
\eta_{\rm O}^{i+1/2,j,k}&=&\frac{\eta_{\rm O}^{i+1,j,k}+\eta_{\rm O}^{i,j,k}}{2},\\
\left(\frac{\eta_{\rm A}}{|\mathbf{B}|^2}\right)^{i+1/2,j,k}&=&\frac{1}{2}\left(\frac{\eta_{\rm A}^{i+1,j,k}}{|\mathbf{B}^{i+1,j,k}|^2}+\frac{\eta_{\rm A}^{i,j,k}}{|\mathbf{B}^{i,j,k}|^2}\right).
\end{eqnarray}
For the source terms in eqs.(\ref{srcb}) and (\ref{srce}), the derivatives at the cell centers are discretized using the following formula and its permutations:
\begin{eqnarray}
(\partial_x B_x)^{i,j,k}&=&\frac{B_x^{i+1,j,k}-B_x^{i-1,j,k}}{2\Delta x}.
\end{eqnarray}

\subsection{A.4. Nested-Grid Boundaries}
At the boundaries between different nested-grid levels, we have to connect solutions in different levels. Because our scheme uses the conservative form, this is straightforwardly implemented using the following procedure.
\begin{enumerate}
\item Project the solution at $t=t^n$ from the coarser grids to the finer grid boundaries using linear interpolation.
\item Calculate fluxes in all the grid levels.
\item Recalculate the fluxes at the level boundaries in the coarser grid using the sum of the fluxes in the finer grid; e.g. $\Delta S F_x^{I+1/2,J,K,l}=\Delta s (F_x^{i+1/2,j,k,l+1}+F_x^{i+1/2,j+1,k,l+1}+F_x^{i+1/2,j,k+1,l+1}+F_x^{i+1/2,j+1,k+1,l+1})$ where $(I,J,K)$ and $(i,j,k)$ are the corresponding cell indices in the coarser and finer grid levels, $l$ is the index of the nested-grid level (larger is finer), and $\Delta S$ and $\Delta s$ are the surface areas of the coarser and finer cells, respectively.
\item Update the solution to $t=t^{n+1}$ in all the grid levels using the corrected fluxes.
\end{enumerate}
Note that no time-interpolation is required because we use shared time stepping.

\subsection{A.5. Time Integration: Super Time Stepping}
As in Paper I, we adopt the Super Time Stepping method \citep[STS,][]{sts,choi09,com11} in order to accelerate the calculation when the time step for the non-ideal MHD part is much smaller than the MHD part. There are two parameters in STS; $\nu$ which is a small positive parameter controlling the stability and efficiency of the scheme, and $N_{\rm STS}$ which is the number of the sub steps used in the scheme. We use $\nu=0.01$ and $N_{\rm STS}=6$ as in Paper I. Acceleration by a factor of $\sim 4$ is achieved with these parameters compared to the first-order explicit scheme.

We use the same time step in all the grid levels. The time step for the non-ideal MHD part is estimated from the standard CFL condition for the diffusion-like equations:
\begin{eqnarray}
\Delta t_{\rm diss}=0.1 \times \min \left(\frac{\Delta x^2}{\eta_{\rm O}+\eta_{\rm A}}\right),
\end{eqnarray}
where 0.1 is a safety factor. When this time step is significantly smaller than that for the MHD part ($\Delta t_{\rm diss} < \Delta t_{\rm MHD} / N_{\rm STS}$), we update the non-ideal MHD part using STS and sub-cycling. If this condition is not satisfied but the time step is still smaller than that for the MHD part ($\Delta t_{\rm MHD} / N_{\rm STS} < \Delta t_{\rm diss} < \Delta t_{\rm MHD}$), the standard forward Euler integrator with sub-cycling is used. If it is larger than the MHD time step ($\Delta t_{\rm MHD} < \Delta t_{\rm diss}$), we update the non-ideal MHD part using the time step for the MHD part.

\subsection{A.6. Tests: Oblique C-shock Problems}
\begin{figure*}[t]
\begin{center}
\includegraphics{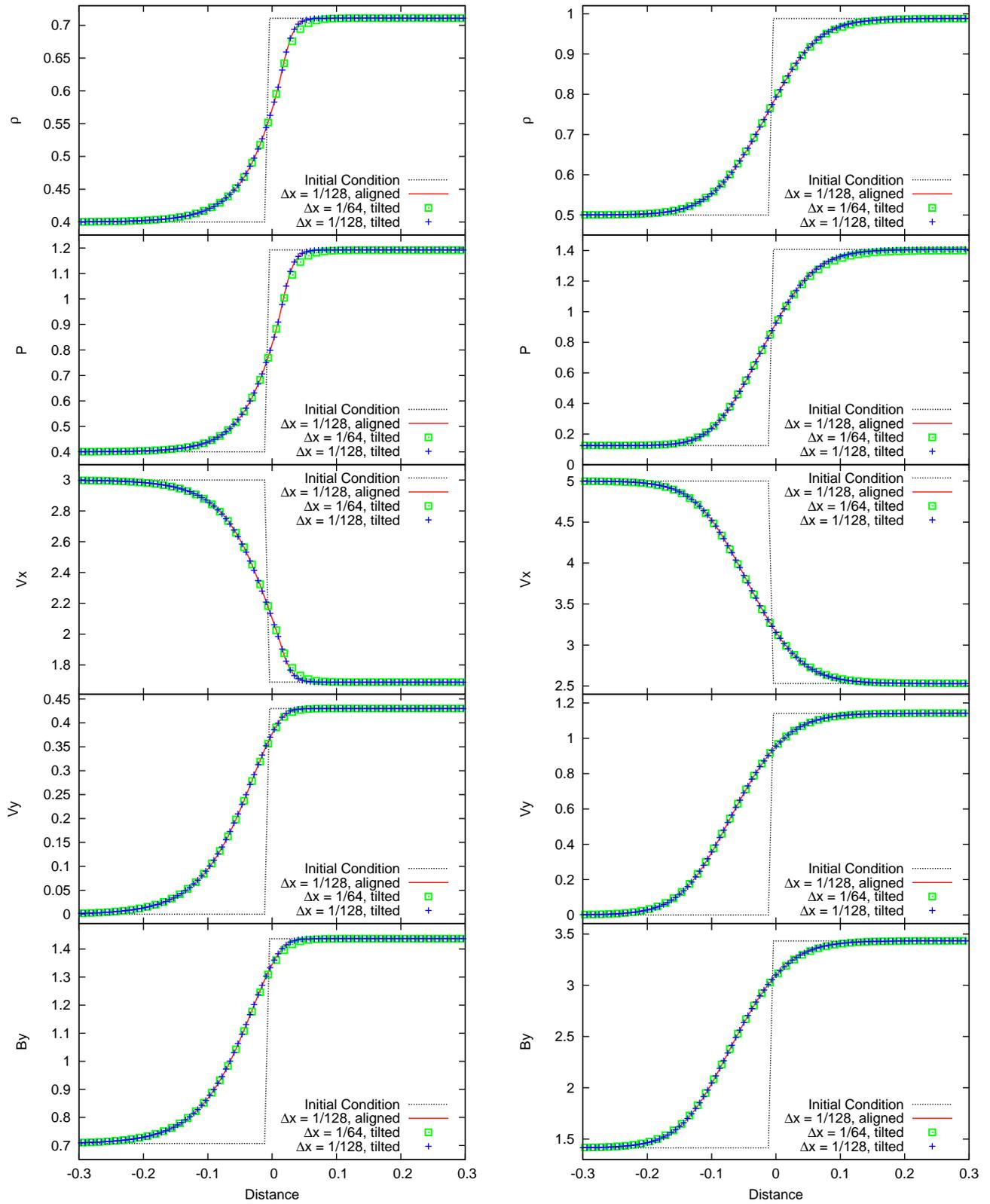}
\caption{The results of the oblique C-shock tests with Ohmic dissipation (left) and ambipolar diffusion (right).}
\label{test}
\end{center}
\end{figure*}
Here we demonstrate the validity of our non-ideal MHD solver based on the non-isothermal oblique C-shock tests \citep{wardle91,ml95,dp08,masson}. These are simple one-dimensional shock-tube tests with the non-ideal MHD effects. 

Although these problems are one-dimensional, we make them two-dimensional tests by adopting the skew-periodic boundary condition. We rotate the shock tube tests by an angle of $\theta$. When $r\equiv \tan\theta$ is a rational number, we can construct the skew-periodic boundary condition which does not introduce any artificial effect. In the $x$-direction, we adopt the fixed boundary condition, which may violate the solenoidal constraint but the ${\rm div}\mathbf{B}$ error is cleaned by the mixed cleaning scheme \citep{dedner} and does not affect the calculations. In the $y$-direction, we copy all the variables from the other side of the computational domain with a shift;
\begin{eqnarray}
v(i,N_y+1)&=&v(i-rN_y,1),
\end{eqnarray}
where $N_y$ is the number of cells in the $y$-direction. If we set $N_y$ so that $rN_y$ is an integer, this boundary condition produces no artificial effect and can be used to run one-dimensional shock tube tests on a two-dimensional grid.

We compare the results between three configurations: fiducial high-resolution aligned models ($\Delta x=1/128$ and $r=0$), high-resolution tilted models ($\Delta x=1/128$, $r=3/4$ and $N_y=4$), and low-resolution tilted models ($\Delta x=1/64$ and $r=3/4$). The rotation angle in the tilted models is $\theta\sim 36.87 \, {\rm deg}$. Each cell is square shaped; $\Delta y=\Delta x$. The resolution in our high-resolution aligned model is corresponding to the highest AMR level in \citet{masson}.

We adopt the same parameter sets as \citet{masson}. For Ohmic dissipation C-shock, the initial left state is $(\rho, v_x, v_y, B_x, B_y, P)=(0.4, 3, 0, \sqrt{2}/2, \sqrt{2}/2, 0.4)$, the right is $(\rho, v_x, v_y, B_x, B_y, P)$ $=(0.71084, 1.68814, 0.4299, \sqrt{2}/2,$ $1.43667, 1.19222)$, and the uniform resistivity $\eta_{\rm O}=0.1$ is used. For ambipolar diffusion, the left is $(\rho, v_x, v_y, B_x, B_y, P)$ $=(0.5, 5, 0, \sqrt{2}, \sqrt{2}, 0.125)$, the right is $(\rho, v_x, v_y, B_x, B_y, P)=(0.9880, 2.5303, 1.1415, \sqrt{2}, 3.4327, 1.4075)$, and the ambipolar diffusion rate is density dependent; $\eta_{\rm A}=1/(75\rho)$. We run the simulations until steady solutions are obtained. In both calculations, the adiabatic index is $\Gamma=5/3$, the CFL number is $0.25$, and the STS time integrator is selected. The results are shown in Figure~\ref{test}. Overall, we successfully obtained good numerical solutions consistent with the semi-analytic solutions. In particular, we found no problem in the multi-dimensional (i.e. tilted) models.

\end{document}